\documentclass{aastex701}

\begin{document}


\title{TeV Gamma-Ray Emission in the DA 495 Region from HAWC
Observations of J1952+2924, J1954+2837, and J1958+2848}

\author{R.~Alfaro}
\affiliation{Instituto de F\'{i}sica, Universidad Nacional Autónoma de México, Ciudad de Mexico, Mexico}
\email{ruben@fisica.unam.mx}

\author{C.~Alvarez}
\affiliation{Universidad Autónoma de Chiapas, Tuxtla Gutiérrez, Chiapas, Mexico}
\email{crabpulsar@hotmail.com}

\author{E.~Anita-Rangel}
\affiliation{Instituto de Astronom\'{i}a, Universidad Nacional Autónoma de México, Ciudad de Mexico, Mexico}
\email{earangel@astro.unam.mx}

\author[orcid=0000-0002-0595-9267]{M.~Araya}
\affiliation{Universidad de Costa Rica, San José 2060, Costa Rica}
\email{miguel.araya@ucr.ac.cr}

\author{J.C.~Arteaga-Velázquez}
\affiliation{Universidad Michoacana de San Nicolás de Hidalgo, Morelia, Mexico}
\email{juan.arteaga@umich.mx}

\author[orcid=0000-0002-4020-4142]{D.~Avila Rojas}
\affiliation{Instituto de Astronom\'{i}a, Universidad Nacional Autónoma de México, Ciudad de Mexico, Mexico}
\email{doavila@astro.unam.mx}

\author[orcid=0000-0002-2084-5049]{H.A.~Ayala Solares}
\affiliation{Temple University, Department of Physics, 1925 N. 12th Street, Philadelphia, PA 19122, USA}
\email{hugo.ayala.solares@temple.edu}

\author[orcid=0000-0002-5529-6780]{R.~Babu}
\affiliation{Department of Physics and Astronomy, Michigan State University, East Lansing, MI, USA}
\email{baburish@msu.edu}

\author{P.~Bangale}
\affiliation{Temple University, Department of Physics, 1925 N. 12th Street, Philadelphia, PA 19122, USA}
\email{priyadarshini.bangale@temple.edu}

\author{A.~Bernal}
\affiliation{Instituto de Astronom\'{i}a, Universidad Nacional Autónoma de México, Ciudad de Mexico, Mexico}
\email{abel@astro.unam.mx}

\author[orcid=0000-0003-2158-2292]{T.~Capistrán}
\affiliation{Instituto Nacional de Astrofísica, Óptica y Electrónica, Puebla, Mexico}
\email{tcapistranc@gmail.com}

\author[orcid=0000-0002-8553-3302]{A.~Carramiñana}
\affiliation{Instituto Nacional de Astrofísica, Óptica y Electrónica, Puebla, Mexico}
\email{alberto@inaoep.mx}

\author[orcid=0000-0002-6144-9122]{S.~Casanova}
\affiliation{Instytut Fizyki Jadrowej im Henryka Niewodniczanskiego Polskiej Akademii Nauk, IFJ-PAN, Krakow, Poland}
\email{sabrinacasanova@gmail.com}

\author[orcid=0000-0002-7607-9582]{U.~Cotti}
\affiliation{Universidad Michoacana de San Nicolás de Hidalgo, Morelia, Mexico}
\email{umberto.cotti@umich.mx}

\author[orcid=0000-0002-1132-871X]{J.~Cotzomi}
\affiliation{Facultad de Ciencias F\'{i}sico Matemáticas, Benemérita Universidad Autónoma de Puebla, Puebla, Mexico}
\email{jcotzomi@yahoo.com.mx}

\author[orcid=0000-0002-7747-754X]{S.~Coutiño de León}
\affiliation{Department of Physics, University of Wisconsin-Madison, Madison, WI, USA}
\email{sara.cdl989@gmail.com}

\author[orcid=0000-0001-9643-4134]{E.~De la Fuente}
\affiliation{Departamento de F\'{i}sica, Centro Universitario de Ciencias Exactase Ingenierias, Universidad de Guadalajara, Guadalajara, Mexico}
\email{edfuente@gmail.com}

\author[orcid=0000-0002-7574-1298]{N.~Di Lalla}
\affiliation{Department of Physics, Stanford University: Stanford, CA 94305–4060, USA}
\email{niccolo.dilalla@stanford.edu}

\author{R.~Diaz Hernandez}
\affiliation{Instituto Nacional de Astrofísica, Óptica y Electrónica, Puebla, Mexico}
\email{dihera77@gmail.com}

\author[orcid=0000-0002-2987-9691]{M.A.~DuVernois}
\affiliation{Department of Physics, University of Wisconsin-Madison, Madison, WI, USA}
\email{duvernois@icecube.wisc.edu}

\author[orcid=0000-0002-0087-0693]{J.C.~Díaz-Vélez}
\affiliation{Department of Physics, University of Wisconsin-Madison, Madison, WI, USA}
\email{juancarlos@icecube.wisc.edu}

\author[orcid=0000-0001-7074-1726]{C.~Espinoza}
\affiliation{Instituto de F\'{i}sica, Universidad Nacional Autónoma de México, Ciudad de Mexico, Mexico}
\email{m.catalina@fisica.unam.mx}

\author[orcid=0000-0002-0173-6453]{N.~Fraija}
\affiliation{Instituto de Astronom\'{i}a, Universidad Nacional Autónoma de México, Ciudad de Mexico, Mexico}
\email{nifraija@astro.unam.mx}

\author{S.~Fraija}
\affiliation{Instituto de Astronom\'{i}a, Universidad Nacional Autónoma de México, Ciudad de Mexico, Mexico}
\email{sarafraija@hotmail.com}

\author{A.~Galván-Gámez}
\affiliation{Instituto de F\'{i}sica, Universidad Nacional Autónoma de México, Ciudad de Mexico, Mexico}
\email{edwin@fisica.unam.mx}

\author[orcid=0000-0002-4188-5584]{J.A.~García-González}
\affiliation{Tecnologico de Monterrey, Escuela de Ingenier\'{i}a y Ciencias, Ave. Eugenio Garza Sada 2501, Monterrey, N.L., Mexico, 64849}
\email{anteus79@tec.mx}

\author[orcid=0000-0003-1122-4168]{F.~Garfias}
\affiliation{Instituto de Astronom\'{i}a, Universidad Nacional Autónoma de México, Ciudad de Mexico, Mexico}
\email{fergar@astro.unam.mx}

\author{N.~Ghosh}
\affiliation{Department of Physics, Michigan Technological University, Houghton, MI, USA}
\email{nghosh1@mtu.edu}

\author{A.~Gonzalez Muñoz}
\affiliation{Instituto de F\'{i}sica, Universidad Nacional Autónoma de México, Ciudad de Mexico, Mexico}
\email{adiv.gonzalez@itoaxaca.edu.mx}

\author[orcid=0000-0002-5209-5641]{M.M.~González}
\affiliation{Instituto de Astronom\'{i}a, Universidad Nacional Autónoma de México, Ciudad de Mexico, Mexico}
\email{magda@astro.unam.mx}

\author{J.A.~González}
\affiliation{Universidad Michoacana de San Nicolás de Hidalgo, Morelia, Mexico}
\email{jose.gonzalez.c@umich.mx}

\author[orcid=0000-0002-9790-1299]{J.A.~Goodman}
\affiliation{Department of Physics, University of Maryland, College Park, MD, USA}
\email{goodman@umd.edu}

\author{J.~Gyeong}
\affiliation{Department of Physics, Sungkyunkwan University, Suwon 16419, South Korea}
\email{kyoungjh1011@naver.com}

\author[orcid=0000-0001-9844-2648]{J.P.~Harding}
\affiliation{Physics Division, Los Alamos National Laboratory, Los Alamos, NM, USA}
\email{jpharding@lanl.gov}

\author[orcid=0000-0001-5169-723X]{I.~Herzog}
\affiliation{Department of Physics and Astronomy, Michigan State University, East Lansing, MI, USA}
\email{herzogia@msu.edu}

\author[orcid=0000-0002-5447-1786]{D.~Huang}
\affiliation{Department of Physics and Astronomy, University of Delaware, Newark, DE, USA}
\email{dezhih@mtu.edu}

\author[orcid=0000-0002-5527-7141]{F.~Hueyotl-Zahuantitla}
\affiliation{Universidad Autónoma de Chiapas, Tuxtla Gutiérrez, Chiapas, Mexico}
\email{filihz@gmail.com}

\author[orcid=0000-0002-3302-7897]{P.~Hüntemeyer}
\affiliation{Department of Physics, Michigan Technological University, Houghton, MI, USA}
\email{petra@mtu.edu}

\author[orcid=0000-0001-5811-5167]{A.~Iriarte}
\affiliation{Instituto de Astronom\'{i}a, Universidad Nacional Autónoma de México, Ciudad de Mexico, Mexico}
\email{airiarte@astro.unam.mx}

\author{S.~Kaufmann}
\affiliation{Universidad Politecnica de Pachuca, Pachuca, Hgo, Mexico}
\email{skaufmann13@googlemail.com}

\author[orcid=0000-0003-4785-0101]{D.~Kieda}
\affiliation{Department of Physics and Astronomy, University of Utah, Salt Lake City, UT, USA}
\email{dave.kieda@utah.edu}

\author{K.~Leavitt}
\affiliation{Department of Physics, Michigan Technological University, Houghton, MI, USA}
\email{kleavitt@mtu.edu}

\author[orcid=0000-0002-2467-5673]{W.H.~Lee}
\affiliation{Instituto de Astronom\'{i}a, Universidad Nacional Autónoma de México, Ciudad de Mexico, Mexico}
\email{wlee@astro.unam.mx}

\author[orcid=0000-0001-5516-4975]{H.~León Vargas}
\affiliation{Instituto de F\'{i}sica, Universidad Nacional Autónoma de México, Ciudad de Mexico, Mexico}
\email{hleonvar@fisica.unam.mx}

\author[orcid=0000-0001-8825-3624]{A.L.~Longinotti}
\affiliation{Instituto de Astronom\'{i}a, Universidad Nacional Autónoma de México, Ciudad de Mexico, Mexico}
\email{alonginotti@astro.unam.mx}

\author[orcid=0000-0003-2810-4867]{G.~Luis-Raya}
\affiliation{Universidad Politecnica de Pachuca, Pachuca, Hgo, Mexico}
\email{gilura6969@hotmail.com}

\author[orcid=0000-0001-8088-400X]{K.~Malone}
\affiliation{Physics Division, Los Alamos National Laboratory, Los Alamos, NM, USA}
\email{kmalone@lanl.gov}

\author[orcid=0000-0001-9052-856X]{O.~Martinez}
\affiliation{Facultad de Ciencias F\'{i}sico Matemáticas, Benemérita Universidad Autónoma de Puebla, Puebla, Mexico}
\email{omartin@fcfm.buap.mx}

\author[orcid=0000-0002-2824-3544]{J.~Martínez-Castro}
\affiliation{Centro de Investigaci\'on en Computaci\'on, Instituto Polit\'ecnico Nacional, M\'exico City, Mexico}
\email{macj@cic.ipn.mx}

\author[orcid=0000-0002-2610-863X]{J.A.~Matthews}
\affiliation{Department of Physics and Astronomy, University of New Mexico, Albuquerque, NM, USA}
\email{johnm@unm.edu}

\author[orcid=0000-0002-8390-9011]{P.~Miranda-Romagnoli}
\affiliation{Universidad Autónoma del Estado de Hidalgo, Pachuca, Mexico}
\email{pa.miranda.r@gmail.com}

\author{P.E.~Mirón-Enriquez}
\affiliation{Instituto de Astronom\'{i}a, Universidad Nacional Autónoma de México, Ciudad de Mexico, Mexico}
\email{pelimi92@gmail.com}

\author[orcid=0000-0001-9361-0147]{J.A.~Morales-Soto}
\affiliation{Universidad Michoacana de San Nicolás de Hidalgo, Morelia, Mexico}
\email{jmoralessg@gmail.com}

\author[orcid=0000-0002-1114-2640]{E.~Moreno}
\affiliation{Facultad de Ciencias F\'{i}sico Matemáticas, Benemérita Universidad Autónoma de Puebla, Puebla, Mexico}
\email{emoreno@fcfm.buap.mx}

\author[orcid=0000-0002-7675-4656]{M.~Mostafá}
\affiliation{School of Science \& Engineering, University of Missouri–Kansas City, Kansas City, MO, USA}
\email{miguel@psu.edu}

\author{M.~Najafi}
\affiliation{Department of Physics, Michigan Technological University, Houghton, MI, USA}
\email{mnajafi@mtu.edu}

\author[orcid=0000-0003-1059-8731]{L.~Nellen}
\affiliation{Instituto de Ciencias Nucleares, Universidad Nacional Autónoma de México, Ciudad de Mexico, Mexico}
\email{lukas@nucleares.unam.mx}

\author[orcid=0000-0002-6859-3944]{M.U.~Nisa}
\affiliation{Department of Physics and Astronomy, Michigan State University, East Lansing, MI, USA}
\email{nisamehr@msu.edu}

\author[orcid=0000-0001-7099-108X]{R.~Noriega-Papaqui}
\affiliation{Universidad Autónoma del Estado de Hidalgo, Pachuca, Mexico}
\email{ropapaqui@gmail.com}

\author[orcid=0009-0009-2481-6921]{M.~Osorio-Archila}
\affiliation{Instituto de Astronom\'{i}a, Universidad Nacional Autónoma de México, Ciudad de Mexico, Mexico}
\email{jmosorio@astro.unam.mx}

\author{E.~Ponce}
\affiliation{Facultad de Ciencias F\'{i}sico Matemáticas, Benemérita Universidad Autónoma de Puebla, Puebla, Mexico}
\email{eponce@fcfm.buap.mx}

\author[orcid=0000-0002-8774-8147]{Y.~Pérez Araujo}
\affiliation{Instituto de F\'{i}sica, Universidad Nacional Autónoma de México, Ciudad de Mexico, Mexico}
\email{yuniorpy@gmail.com}

\author[orcid=0000-0001-5998-4938]{E.G.~Pérez-Pérez}
\affiliation{Universidad Politecnica de Pachuca, Pachuca, Hgo, Mexico}
\email{egperezp@yahoo.com.mx}

\author[orcid=0000-0002-6524-9769]{C.D.~Rho}
\affiliation{Department of Physics, Sungkyunkwan University, Suwon 16419, South Korea}
\email{no397@naver.com}

\author[orcid=0000-0003-1327-0838]{D.~Rosa-González}
\affiliation{Instituto Nacional de Astrofísica, Óptica y Electrónica, Puebla, Mexico}
\email{danrosa@inaoep.mx}

\author[orcid=0000-0002-4204-5026]{M.~Roth}
\affiliation{Physics Division, Los Alamos National Laboratory, Los Alamos, NM, USA}
\email{mattroth@lanl.gov}

\author[orcid=0000-0001-6079-2722]{A.~Sandoval}
\affiliation{Instituto de F\'{i}sica, Universidad Nacional Autónoma de México, Ciudad de Mexico, Mexico}
\email{asandoval@fisica.unam.mx}

\author{J.~Serna-Franco}
\affiliation{Instituto de F\'{i}sica, Universidad Nacional Autónoma de México, Ciudad de Mexico, Mexico}
\email{j_serna@ciencias.unam.mx}

\author{M.~Shin}
\affiliation{Department of Physics, Sungkyunkwan University, Suwon 16419, South Korea}
\email{minjishin23@gmail.com}

\author[orcid=0000-0002-1012-0431]{A.J.~Smith}
\affiliation{Department of Physics, University of Maryland, College Park, MD, USA}
\email{asmith8@umd.edu}

\author{Y.~Son}
\affiliation{Department of Physics, Sungkyunkwan University, Suwon 16419, South Korea}
\email{youngwan.son@cern.ch}

\author[orcid=0000-0002-1492-0380]{R.W.~Springer}
\affiliation{Department of Physics and Astronomy, University of Utah, Salt Lake City, UT, USA}
\email{wayne.springer@utah.edu}

\author{O.~Tibolla}
\affiliation{Universidad Politecnica de Pachuca, Pachuca, Hgo, Mexico}
\email{omar.tibolla@gmail.com}

\author[orcid=0000-0002-1689-3945]{I.~Torres}
\affiliation{Instituto Nacional de Astrofísica, Óptica y Electrónica, Puebla, Mexico}
\email{ibrahim.torres23@gmail.com}

\author[orcid=0000-0002-7102-3352]{R.~Torres-Escobedo}
\affiliation{Tsung-Dao Lee Institute \& School of Physics and Astronomy, Shanghai Jiao Tong University, 800 Dongchuan Rd, Shanghai, SH 200240, China}
\email{torresramiro350@sjtu.edu.cn}

\author[orcid=0000-0001-6876-2800]{L.~Villaseñor}
\affiliation{Facultad de Ciencias F\'{i}sico Matemáticas, Benemérita Universidad Autónoma de Puebla, Puebla, Mexico}
\email{lvillasen@gmail.com}

\author[orcid=0000-0001-6798-353X]{X.~Wang}
\affiliation{Department of Physics, Missouri University of Science and Technology, Rolla, MO, USA}
\email{xiaojiewang@mst.edu}

\author{Z.~Wang}
\affiliation{Department of Physics, Missouri University of Science and Technology, Rolla, MO, USA}
\email{zhen@umd.edu}

\author[orcid=0000-0003-2141-3413]{I.J.~Watson}
\affiliation{University of Seoul, Seoul, South Korea}
\email{ian.james.watson@cern.ch}

\author[orcid=0009-0005-7243-1402]{H.~Wu}
\affiliation{Department of Physics, University of Wisconsin-Madison, Madison, WI, USA}
\email{hwu298@wisc.edu}

\author[orcid=0009-0006-3520-3993]{S.~Yu}
\affiliation{Department of Physics, Pennsylvania State University, University Park, PA, USA}
\email[show]{sjy5345@psu.edu} 
\correspondingauthor{Seonghyeon Yu}

\author[orcid=0000-0002-9307-0133]{S.~Yun-Cárcamo}
\affiliation{Department of Physics, Drexel University, 3141 Chestnut Street, Philadelphia, PA 19104, USA}
\email{yunsoh@umd.edu}

\author{X.~Zhang}
\affiliation{Instytut Fizyki Jadrowej im Henryka Niewodniczanskiego Polskiej Akademii Nauk, IFJ-PAN, Krakow, Poland}
\email{xiyingzhangxyz@gmail.com}

\author[orcid=0000-0003-0513-3841]{H.~Zhou}
\affiliation{Tsung-Dao Lee Institute \& School of Physics and Astronomy, Shanghai Jiao Tong University, 800 Dongchuan Rd, Shanghai, SH 200240, China}
\email{hao_zhou@sjtu.edu.cn}

\author[orcid=0000-0002-8528-9573]{C.~de León}
\affiliation{Universidad Michoacana de San Nicolás de Hidalgo, Morelia, Mexico}
\email{cederik.de.leon@umich.mx}








\collaboration{all}{The HAWC collaboration}


\begin{abstract}

\vspace{0.5em}

Previous observations of the DA 495 region, which contains the pulsar wind nebula (PWN) DA 495 and other very-high-energy (VHE; $>100~\textrm{GeV}$) gamma-ray emitters, showed inconsistent results across different VHE experiments, including differences in the measured flux of DA 495 and in the number of reported sources. Our new dedicated analysis using 3071 days of HAWC data resolves these discrepancies. We identify three extended sources, HAWC J1952+2924 (DA 495), HAWC J1954+2837, and HAWC J1958+2848, all with curved spectra, and find a hint of a fourth source. For DA 495, a multi-wavelength analysis, including our results, radio, X-ray, and other gamma-ray data, reveals that its extended emission is best described by leptonic interactions with an inferred magnetic field of $B\sim10~\mu\textrm{G}$ and that hadronic components are not required. On the other hand, we interpret the other two sources, which have no confirmed counterparts, as TeV halos associated with nearby pulsars, PSR J1954+2836 and PSR J1958+2846, respectively. These results reinforce the interpretation of DA 495 as a leptonic-dominated PWN and add two candidates to the growing population of TeV halos.

\end{abstract}



\section{Introduction}
Observations of very-high-energy (VHE; $>$ 100 GeV) gamma rays allow us to probe extreme astrophysical environments, such as supernova remnants (SNRs) and pulsar wind nebulae (PWNs), and play an essential role in understanding the origin of cosmic rays. However, resolving VHE gamma-ray sources is a major challenge when their projected angular separations are comparable to the angular resolution of the observatories. This difficulty increases for sources near the Galactic plane due to the Galactic diffuse emission (GDE) generated by interactions of cosmic rays within the Galaxy. The DA 495 region, where the PWN 495 lies near several VHE gamma-ray sources, is one such example.  

DA 495 was first detected as a point-like source by the Dominion Astrophysical (DA) survey in the radio band \citep{1968AJ.....73..135G}. However, subsequent radio observations revealed an extended morphology \citep{1973A&A....26..237W} and a centrally peaked emissivity distribution declining with increasing radius \citep{1983AJ.....88.1810L}, suggesting a center-filled SNR structure. In contrast, an alternative interpretation proposed that DA 495 may be an SNR with an exceptionally thick shell \citep{1989JApA...10..161V}. The PWN interpretation of DA 495 was confirmed by its first X-ray detection using ROSAT and ASCA observations \citep{2004ApJ...610L.101A}, identifying a compact object associated with the nebula. Later, the most recent radio analysis to date, \citet{2008ApJ...687..516K}, established it as an evolving PWN with a pronounced spectral break, and further X-ray observations using Chandra and XMM-Newton confirmed the non-thermal region surrounding a central neutron star \citep{2008ApJ...687..505A, 2015MNRAS.453.2241K}.

The VHE gamma-ray emission from DA 495 was first reported by the High Altitude Water Cherenkov (HAWC) collaboration in its 2HWC catalog as 2HWC J1953+294 \citep{2017ApJ...843...40A}. Following this detection, the Fermi Large Area Telescope (Fermi-LAT), the Very Energetic Radiation Imaging Telescope Array System (VERITAS), and HAWC conducted a collaborative study of the 14 non-associated sources in the 2HWC catalog across energy ranges from GeV to TeV \citep{2018ApJ...866...24A}. In this follow-up analysis, VERITAS significantly detected DA 495, confirming it as a VHE gamma-ray source. However, the nearby source 2HWC J1955+285 was not detected at the time. Moreover, the differential flux of DA 495 at 1 TeV measured by HAWC was about 7 times higher than that measured by VERITAS. This discrepancy was suggested to arise from unresolved gamma-ray emission near DA 495, leading to an overestimation of DA 495's flux in the HAWC measurement. 

In the 3HWC catalog \citep{2020ApJ...905...76A}, HAWC reported three sources in the DA 495 region, partially supporting the unresolved-source scenario. However, the flux of DA 495 measured by HAWC remained higher than that measured by VERITAS by a factor of 3. The same three sources were reported in the DA 495 region in the 4HWC catalog \citep{2026arXiv260200263A}, but the flux discrepancy was not resolved. On the other hand, the Large High Altitude Air Shower Observatory (LHAASO) collaboration reported four sources in the same region in its first catalog \citep{2024ApJS..271...25C}. Three of them, including DA 495, correspond to the HAWC sources, and the LHAASO collaboration further explored DA 495, reporting gamma-ray emission extending beyond 100 TeV and an extended X-ray morphology \citep{2026arXiv260111071T}. In contrast, the fourth source did not match any HAWC sources and is characterized by a large spatial extension, measured as $0.99^\circ$ by the LHAASO Water Cherenkov Detector Array (WCDA), and $0.78^\circ$ by the LHAASO kilometer square array (KM2A).

As different VHE gamma-ray experiments reported inconsistent results, we perform a new dedicated analysis of this region to investigate the observed discrepancies and examine the identified sources. The paper is structured as follows. In Section \ref{sec:HAWC}, we describe the HAWC Observatory and the data set we use in this work. In Section \ref{sec:DataAnalysis}, we present our analysis procedure, and provide the modeling of the sources in Section \ref{sec:SEDModeling}. We then discuss the results and their physical implications in Section \ref{sec:Discussion}. Finally, we conclude our paper in Section \ref{sec:Conclusion}.

\section{HAWC}
\label{sec:HAWC}

\subsection{The HAWC Observatory}
The High-Altitude Water Cherenkov (HAWC) Observatory, located at an altitude of $4100~\mathrm{m}$ above sea level in Sierra Negra, is a particle detector array that detects gamma rays from several hundred GeV to a few hundred TeV through secondary particles from extensive air showers (EAS) produced by primary particles. It comprises 300 water tanks, each with a diameter of $7.3~\mathrm{m}$ and a height of $4.5~\mathrm{m}$, containing approximately $200,000~\mathrm{L}$ of purified water. Three 8-inch and one 10-inch photomultiplier tubes (PMTs) are anchored at the bottom of each tank, and the 8-inch PMTs form an equilateral triangle $1.83~\mathrm{m}$ apart from the center, where the 10-inch PMT is located. Due to its high duty cycle of $95\%$ and wide instantaneous field of view of $2~\mathrm{sr}$, the observatory daily covers approximately two-thirds of the sky, \citep{2023NIMPA105268253A}. 

\subsection{HAWC Data Set}
We use 3071 days of data from the fifth HAWC reconstruction pass \citep{2024ApJ...972..144A}, which is currently the most recent version, with the up-to-date gamma-hadron separation based on machine learning algorithms \citep{2025ApJ...992..156A}. We employ two different schemes. In the morphological analysis and the spectral model selection, we use the $f_{hit}$ scheme, which classifies events based on the fraction of hit PMTs in the detector array. Since events with reconstructed EAS cores both inside and outside the detector array are included, this scheme allows us to take advantage of higher statistics. We then repeat the spectral model selection as a consistency check and perform spectral fitting using an energy-based binning scheme optimized for reconstructing the energy of primary gamma rays. While only events with reconstructed cores inside the detector array are analyzed, this scheme enables us to determine flux points for a given spectral model. Two independent estimators are available for the energy reconstruction: the ground parameter (GP) and neural network (NN) methods. We use the GP estimator, which reconstructs the energy using the effective charge at a fixed distance from the shower axis. A detailed description of the two estimators can be found in \citet{2019ApJ...881..134A}.

\section{Data Analysis and Results}
\label{sec:DataAnalysis}

\subsection{Analysis Framework}
We analyze a $4^\circ\times4^\circ$ region centered at (R.A., Decl.) $=$ ($299^\circ$, $29^\circ$), while considering a $10^\circ\times10^\circ$ domain centered at the same position to account for possible source extensions during the modeling process. In this analysis, the HAWC Accelerated Likelihood (HAL) plugin \citep{2022icrc.confE.828H} within the Multi-Mission Maximum Likelihood (3ML) framework \citep{2015arXiv150708343V} is used to evaluate models. When comparing two models, model $i$ and $j$, we calculate the test statistic (TS) from their maximum likelihood values, $\hat{L}_i$ and $\hat{L}_j$, as
\begin{equation}
\mathrm{TS}=2\ln{\frac{\hat{L}_i}{\hat{L}_j}}.
\label{eq:TSCalculation}
\end{equation}
Here, model $j$ is taken as the null hypothesis, while model $i$ represents the alternative hypothesis. Under the conditions of Wilks' theorem \citep{Wilks}, which is valid for HAWC data \citep{2017ApJ...843...40A}, the TS follows a $\chi^2$ distribution with degrees of freedom equal to the difference in the number of free parameters between the two models. Therefore, when this difference is 1, the pretrial significance $\sigma$ can be approximated as $\sqrt{\mathrm{TS}}$. Based on this principle, we construct significance maps by evaluating the TS at each pixel, where $\hat{L}_i$ represents the maximum likelihood of the signal+background model and $\hat{L}_j$ is that of the background-only model, assuming a Poisson likelihood. Similarly, we define the $\mathrm{TS}$ of an individual source for a given model, $\mathrm{TS}_{\mathrm{source}}$, where the alternative hypothesis corresponds to the maximum likelihood of the full model and the null hypothesis to that of the model without the specific source.

\subsection{Galactic Diffuse Emission}
Since the DA 495 region is close to the Galactic plane ($|b|<2^\circ$), modeling the GDE is crucial for resolving individual sources. To model the GDE, we generate templates using High-Energy Radiative Messengers (HERMES) \citep{2021A&A...653A..18D}, incorporating gamma-ray emission from pion decay and inverse Compton scattering produced by interactions of cosmic rays with the interstellar medium (ISM) and ambient radiation fields. When the templates are generated, the cosmic-ray distribution is assumed to be spatially uniform and isotropic, potentially leading to an underestimation of the diffuse emission. Therefore, we allow the normalization of the GDE templates, $k_{GDE}$, to vary in both the morphological and spectral analyses. 


\subsection{Morphological Analysis}
\label{sec:MorphAnalysis}
\begin{figure}[t]
\centering
\includegraphics[width=0.6\columnwidth]{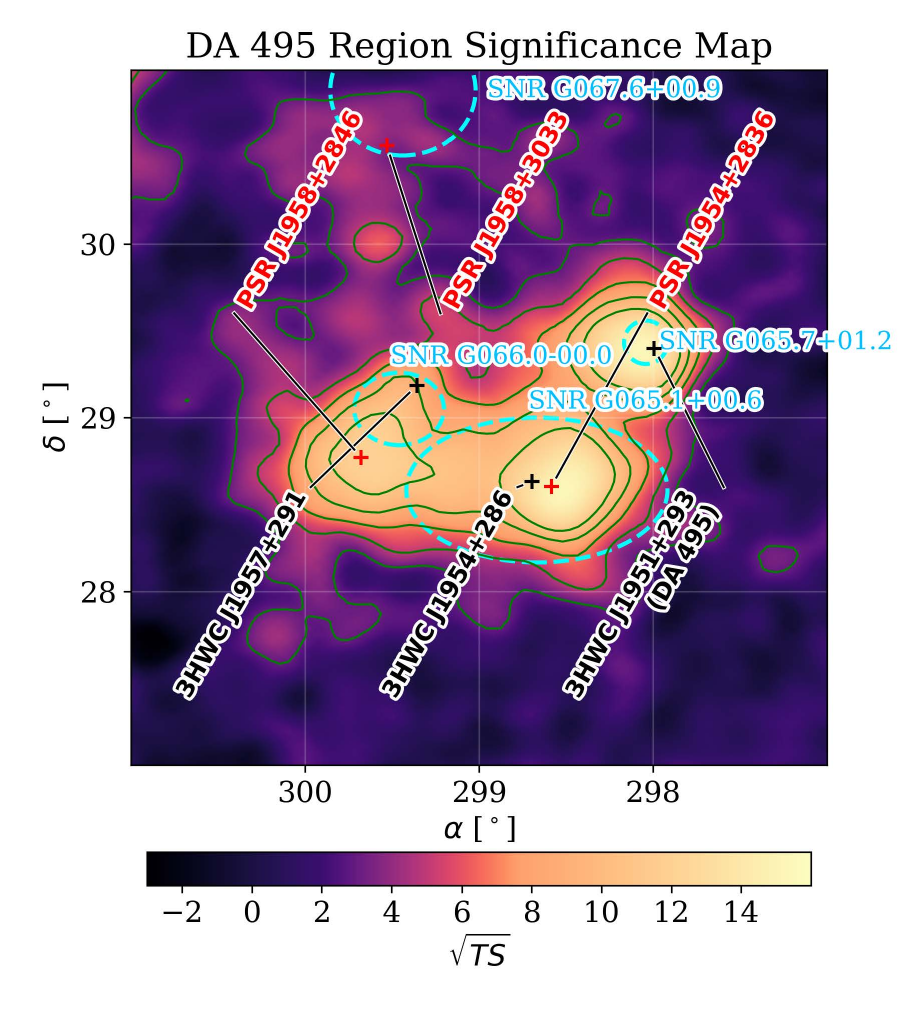}
\caption{Significance map of the DA 495 region. The green contours indicate significance levels of 3, 5, 7, 9, and 11$\sigma$. The black crosses show the 3HWC positions, the red crosses represent the locations of pulsars, and the cyan contours mark supernova remnants identified in radio observations. A spectral index of 2.4 is assumed.}
\label{fig:SignificanceMap}
\end{figure}

\begin{figure*}[t!]
\centering
\gridline{
    \fig{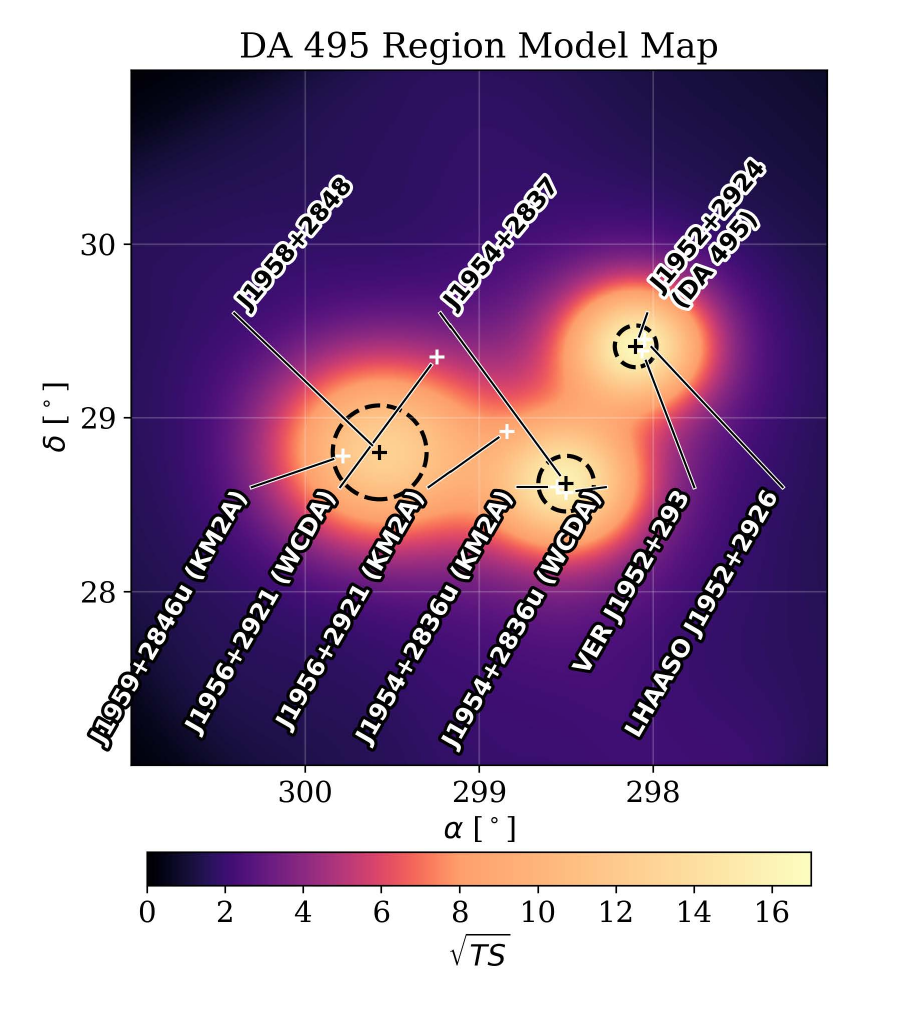}{0.49\textwidth}{(a)}
    \fig{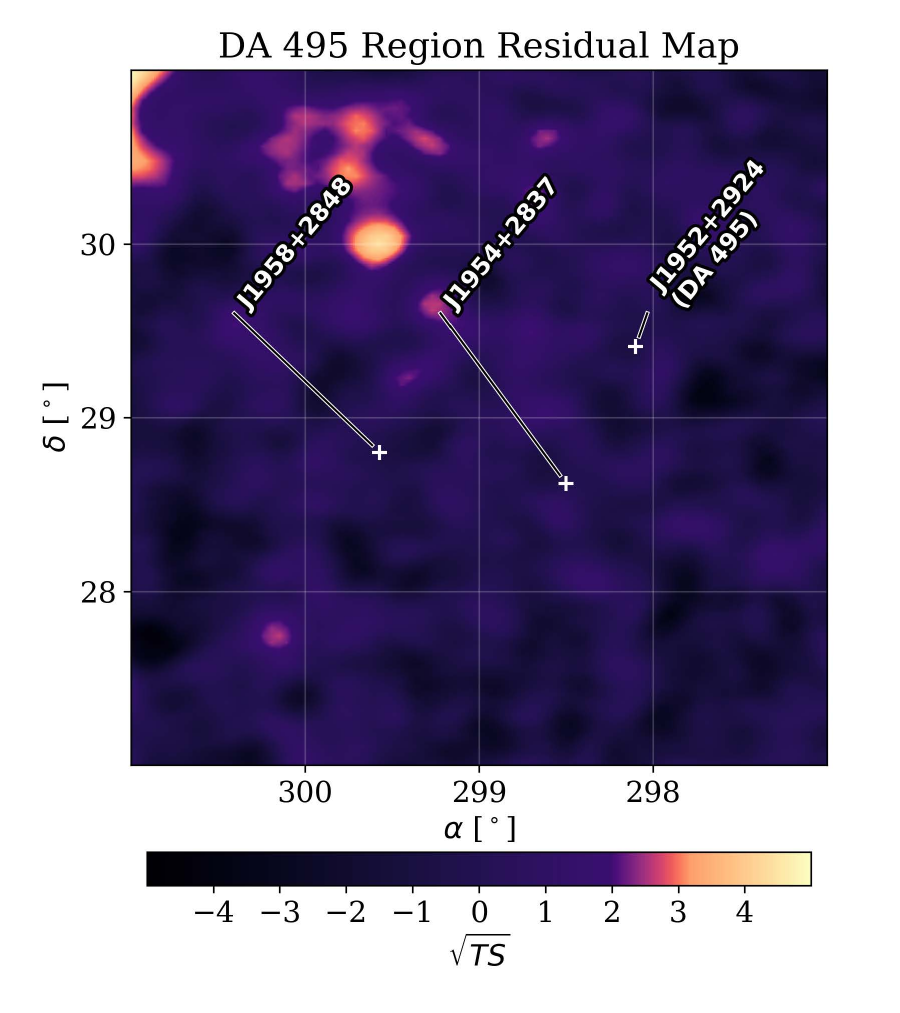}{0.49\textwidth}{(b)}
}
\caption{
(a) Model map of J1952, J1954, and J1958. The black crosses with the labels above them show the new morphological analysis results, whereas the white crosses with the labels below them present results from other experiments. For labels from LHAASO, `LHAASO' denotes joint-fit results, whereas `WCDA' and `KM2A' denote sources without joint fits. The black dashed contours indicate the spatial extensions $\sigma_s$ of the sources.
(b) Residual map without the three identified extended sources and the GDE. The white crosses with the labels above them show the new morphological analysis results. A bright spot with $\sqrt{TS}\sim4$ is visible near (R.A., Decl.) = $(299.6^\circ, 30.0^\circ)$, which may indicate a fourth source. A spectral index of 2.4 is assumed in both maps.
}
\label{fig:ModelResidual3ES}
\end{figure*}

\begin{figure}[t]
\centering
\includegraphics[width=0.7\columnwidth]{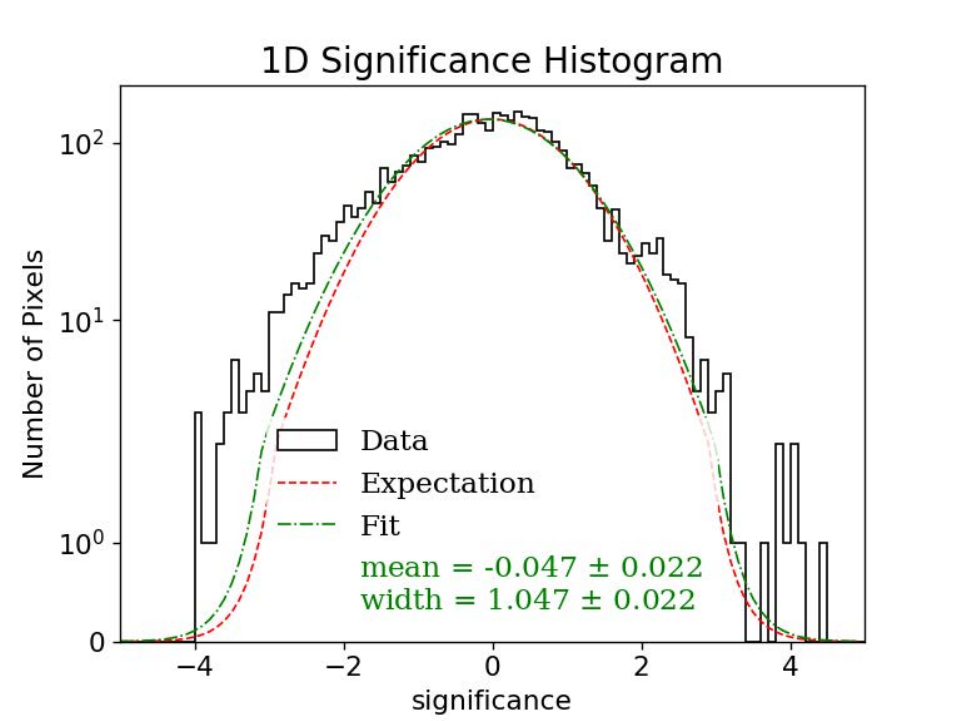}
\caption{1D significance histogram of the residual map in Figure \ref{fig:ModelResidual3ES}. A circular region of $2^\circ$ radius centered at $(\mathrm{R.A.}, \mathrm{Decl.})=(299^\circ, 29^\circ)$ is considered. The additional excess up to a $4 \sigma$ level comes from the bright spot near $(\mathrm{R.A.}, \mathrm{Decl.})=(299.6^\circ, 30.0^\circ)$, suggesting a potential fourth source. On the other hand, the negative-significance tail originates from low-background regions where the estimated background can be negative, for which the Poisson likelihood is not defined and a Gaussian likelihood is used instead.}
\label{fig:1DResidual3ES}
\end{figure}

In the morphological analysis, we perform a blind search to determine the number of sources, their positions, and spatial extensions. For the extension test, we adopt a symmetric 2D Gaussian model on the sphere described as
\begin{equation}
f(x,y)=\frac{A}{2 \pi \sigma_{s}^2}\exp\left(-\frac{(x-x_0)^2+(y-y_0)^2}{2\sigma_s^2}\right)
\label {eq:2DGaussian}
\end{equation}
for a given position (R.A., Decl.)=$(x_0, y_0)$, where $A$ is a normalization constant, and $\sigma_s$ refers to the spatial extension. The blind search starts by adding a point source, which is accepted if the TS increases by more than 25. The added source is classified as an extended source if the extended-source test yields $\Delta\mathrm{TS}>16$. Whenever a new source is added, the extended-source test is performed again for all point sources one by one, allowing previously tested point sources to be classified as extended. This procedure is repeated until no additional point source increases the TS by more than 25. When the search terminates, we additionally test extended morphology for the last sub-threshold source candidate, since a highly extended source may not exceed the point-source detection threshold. The spectrum of a source is modeled as a power-law (PL), which is given by 
\begin{equation}
\frac{dN}{dE}=K\left(\frac{E}{E_0}\right)^{\alpha},
\label{eq:power-law}
\end{equation}
with the pivot energy fixed at $E_0=7~\mathrm{TeV}$. All parameters, including positions, spatial extensions, $K$, and $\alpha$, are allowed to vary, except for the fixed pivot energy. Based on the search algorithm, we identify three extended sources, HAWC J1952+2924 (DA 495), HAWC J1954+2837, and HAWC J1958+2848 (hereafter J1952, J1954, and J1958), with $k_{GDE}=4.9\pm0.5$. Figure~\ref{fig:SignificanceMap} shows the significance map of the DA 495 region, and Figure~\ref{fig:ModelResidual3ES} shows the corresponding model and residual maps. The results are summarized in Table~\ref{tab:MorpResults}. Although a gamma-ray excess remains near (R.A., Decl.) = $(299.6^\circ, 30.0^\circ)$ at a $4\sigma$ level after subtracting the three identified sources and the GDE, as shown in Figures \ref{fig:ModelResidual3ES} and \ref{fig:1DResidual3ES}, adding a point or an extended source at this location does not increase the TS sufficiently. Nevertheless, the resulting TS values are close to the detection threshold. Since the point source and extended source fits have different degrees of freedom, we compare their significances instead of TS values, finding that the extended source model yields a higher significance $(3.97\sigma)$ compared to the point source model $(3.38\sigma)$.


\begin{deluxetable*}{cccccc}
\tablewidth{0pt}
\tablecaption{Morphological analysis results}
\label{tab:MorpResults}
\tablehead{
\colhead{Source} & \colhead{R.A. (deg)} & \colhead{Decl. (deg)} & \colhead{Extension (deg)} & \colhead{$\mathrm{TS}_{\textrm{source}}$} 
}
\startdata
J1952 &
$298.10\pm0.02\pm0.02$& $29.41\pm0.02\pm0.01$ & $0.12\pm0.02\pm0.01$ & $180$ \\
J1954 &
$298.50\pm0.03\pm0.01$ & $28.62\pm0.03\pm0.02$ & $0.16\pm0.03\pm0.01$ & $139$ \\
J1958 &
$299.57\pm0.05\pm0.03$ & $28.80\pm0.04\pm0.04$ & $0.27\pm0.03\pm0.03$ & $165$ \\
\enddata
\tablecomments{The first set of uncertainties is statistical, and the second is systematic. A description of systematic uncertainties is provided in Section \ref{sec:SystematicUncertainties}.}
\end{deluxetable*}


We additionally perform the morphological analysis with $k_{GDE}$ fixed to 1 to assess the impact of diffuse emission on the source search. In this case, we identify four extended sources. Although the number of identified sources differs from the case where $k_{\mathrm{GDE}}$ is allowed to vary, the results from the three sources, J1952, J1954, and J1958, are consistent, demonstrating the robustness of their modeling across different diffuse emission normalizations. The minor differences arising from different diffuse emission models are treated as systematic uncertainties as described in Section \ref{sec:SystematicUncertainties}.

The fourth source is characterized by a large spatial extension of $1.15^\circ$, located at (R.A., Decl.) = $(299.21^\circ, 29.80^\circ)$. Although this source exceeds our detection threshold, we do not claim it as a distinct source because it relies heavily on the normalization of the diffuse emission model. Therefore, we conduct the spectral analysis only for J1952, J1954, and J1958. Additionally, based on our results with varying and fixed $k_{GDE}$, we further discuss this potential extended fourth source in Section \ref{subsec:FourthSource}.

\begin{figure}[t]
\centering
\includegraphics[width=0.6\columnwidth]{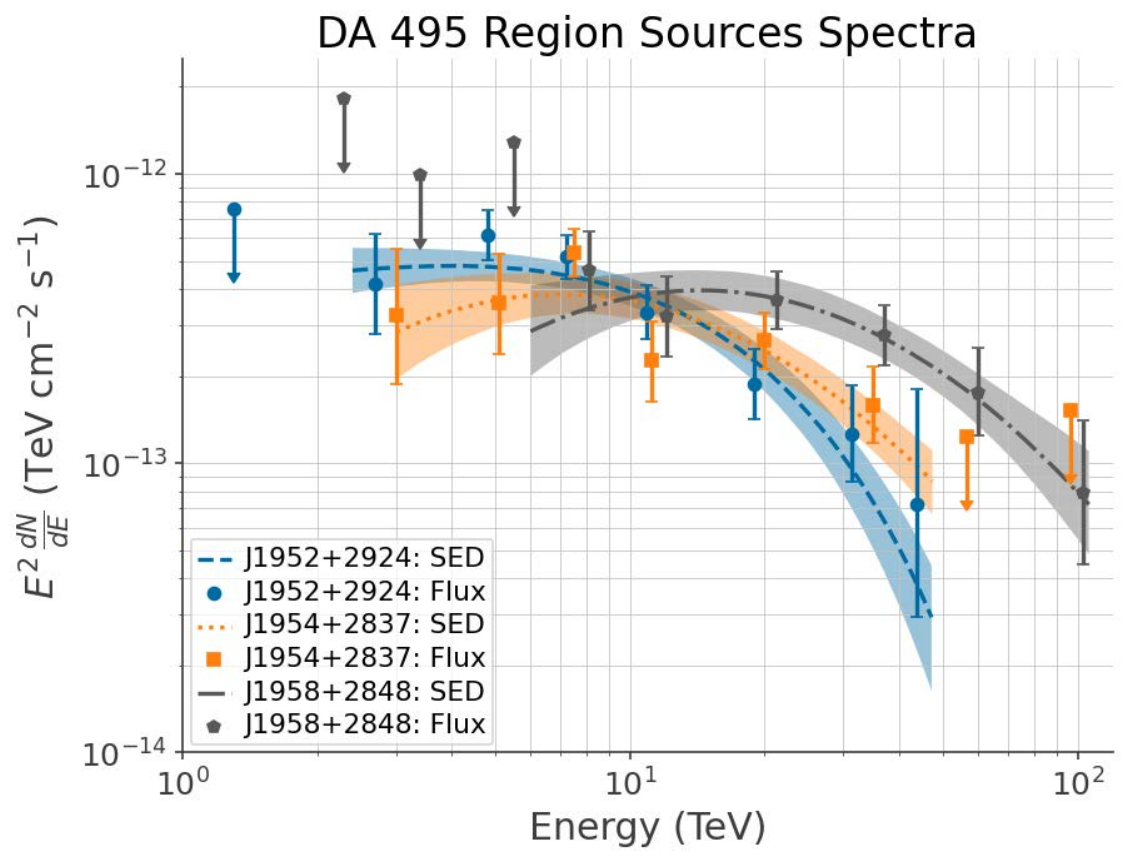}
\caption{SEDs of J1952, J1954, and J1958. They are best described by PLEC, LP, and LP models, respectively. Only statistical uncertainties are considered in the errors and error bands. The flux points are provided in Appendix \ref{app:fluxpoints}.}
\label{fig:Spectra}
\end{figure}

\subsection{Spectral Analysis}
\label{sec:SpecAnalysis}
We perform the spectral analysis after fixing the positions of the sources identified in the morphological analysis, considering the PL (Eq. (\ref{eq:power-law})), log-parabola (LP; Eq. (\ref{eq:log-parabola})), and power-law with an exponential cutoff (PLEC; Eq. (\ref{eq:power-law cutoff})) models, defined as: 
\begin{equation}
\frac{dN}{dE}=K\left(\frac{E}{E_0}\right)^{\alpha-\beta\ln\left(\frac{E}{E_0}\right)},
\label{eq:log-parabola}
\end{equation}
\begin{equation}
\frac{dN}{dE}=K\left(\frac{E}{E_0}\right)^{\alpha}\exp\left(-\frac{E}{E_c}\right).
\label{eq:power-law cutoff}
\end{equation}
We test all possible 27 combinations (three spectral models for each source) and select the model that yields the minimum Akaike information criterion (AIC), calculated as
\begin{equation}
\mathrm{AIC}=-2\ln{L}+2k,
\label{eq:AIC}
\end{equation}
where $L$ is the likelihood value and $k$ is the number of free parameters \citep{1974ITAC...19..716A}. The best spectral models are PLEC for J1952 and LP for both J1954 and J1958. We note that the LP and PLEC models are not significantly favored over each other for all three sources ($\Delta\mathrm{AIC} < 2$), but compared to the PL model, J1958 shows $\Delta\mathrm{AIC}\sim4$, whereas J1952 and J1954 show $\Delta\mathrm{AIC}\sim12$ and 16, respectively, indicating a stronger statistical preference for curved spectra. Figure \ref{fig:Spectra} presents the spectral energy distributions (SEDs) of the sources. After selecting their best spectral models, we determine their energy ranges where they are significantly detected. Specifically, we first identify the lower and upper energies at which the likelihood is maximized. We then obtain the energies $E_{min}$ and $E_{max}$ where the $\ln{L}$ value decreases by $0.5$ relative to the maximum, corresponding to the $1\sigma$ confidence interval. As a result, the final energy ranges are $[E_{min},~E_{max}] = [2.4,~47.2]$, $[3.0,~47.4]$, and $[6.0,~106.0]~\mathrm{TeV}$ for J1952, J1954, and J1958, respectively. Lastly, we choose the optimal set of pivot energies $E_0=(3,20,37)~\mathrm{TeV}$, which minimizes the correlation between the free parameters. The spectral results are summarized in Table \ref{tab:SpecResults}.

\begin{table}[htb!]
\centering
\caption{Spectral analysis results}
\label{tab:SpecResults}
\begin{tabular}{lll}
\hline \hline
Source & Parameter & Best Fit \\
\hline
J1952 
 &$K~$& $6.83\,^{+1.18}_{-0.97}\,^{+0.22}_{-1.19}\times10^{-14}$ \\
 &$\alpha$& $-1.66\pm0.20\,^{+0.09}_{-0.16}$ \\
 &$E_c$& $12.0\,^{+3.1}_{-2.4}\,^{+3.6}_{-1.4}$ \\
\hline
J1954 
& $K$       
& $6.15\,^{+1.01}_{-0.88}\,^{+0.74}_{-0.66}\times10^{-16}$ \\
& $\alpha$& $-2.86\pm0.18\,^{+0.12}_{-0.03}$ \\
& $\beta$& $0.41\pm0.13\,^{+0.02}_{-0.06}$ \\
\hline
J1958 
 & $K$    
 & $1.97\,^{+0.36}_{-0.30}\,^{+0.28}_{-0.28}\times10^{-16}$ \\
& $\alpha$& $-2.80\pm0.23\,^{+0.19}_{-0.04}$ \\
& $\beta$& $0.42\pm0.18\,^{+0.03}_{-0.07}$ \\
\hline
\end{tabular}
\tablecomments{Spectral parameters are obtained with fixed pivot energies of $E_0=3, 20, \textrm{and}~37~\mathrm{TeV}$ for J1952, J1954, and J1958, respectively. The units of $K$ and $E_c$ are $\mathrm{TeV}^{-1}~\mathrm{cm}^{-2}~\mathrm{s}^{-1}$ and $\mathrm{TeV}$. The first set of uncertainties is statistical, and the second is systematic. A description of systematic uncertainties is provided in Section \ref{sec:SystematicUncertainties}.}
\end{table}

\subsection{Systematic Uncertainties}
\label{sec:SystematicUncertainties}
To evaluate systematic uncertainties, we consider the modeling of GDE and variations in different detector properties used in the simulation, including photoelectron pulse width, charge calibration uncertainties, PMT thresholds, and run-dependent detector conditions. Further details on these detector properties can be found in \citet{2019ApJ...881..134A}. We estimate systematic uncertainties due to GDE modeling from the difference in the fit parameters between the case where $k_{GDE}$ is left free and the case where $k_{GDE}$ is fixed to 1. Similarly, we estimate them from the detector properties by computing the shifts in the fit parameters under different detector conditions. Since these contributions do not show correlations, as described in \citet{2019ApJ...881..134A}, we combine the corresponding uncertainties in quadrature to obtain a total systematic uncertainty for each parameter. Regarding the positions and spatial extensions of J1952, J1954, and J1958, we find that these detector-related effects have a negligible impact on the uncertainties, so we estimate the systematic uncertainties solely from the GDE modeling. On the other hand, the spectral parameters are influenced by both the modeling of GDE and the detector properties. Tables \ref{tab:MorpResults} and \ref{tab:SpecResults} include the systematic uncertainties of the fit parameters.

\section{SED Modeling}
\label{sec:SEDModeling}

\subsection{HAWC J1952+2924 (DA 495)}
\label{sec:DA495Modeling}
Since DA 495 has been detected across multiple wavelengths in addition to the VHE gamma-ray band, we perform a multi-wavelength analysis using radio data from 83 MHz to 10 GHz \citep{2008ApJ...687..516K}, X-ray upper limit constraints from 2 to 10 keV \citep{2019ApJ...878..126C}, Fermi-LAT data in the GeV energy band analyzed from \citet{2026ApJ...997..367T}, VERITAS measurements from 100 GeV to 1 TeV \citep{2018ApJ...866...24A}, our HAWC results 
from a few TeV to the multi-TeV regime, and LHAASO KM2A measurements from several tens of TeV up to 2 PeV \citep{2026arXiv260111071T}. We note that we do not include the LHAASO WCDA results \citep{2026arXiv260111071T} in the multi-wavelength fit because there is a significant overlap between the WCDA and HAWC energy ranges, and the WCDA measurements are consistent with our results (See Figure \ref{fig:J1952Comparison}). Therefore, including the WCDA data would primarily add statistical weight to the low TeV-energy regime instead of providing new constraints on the broadband spectral shape. Additionally, we do not include the X-ray spectrum reported by \citet{2026arXiv260111071T}. The reported X-ray emission was obtained by analyzing the same archival data used in \citet{2019ApJ...878..126C} with a different background treatment. Since both results use the same archival data, we adopt the results in \citet{2019ApJ...878..126C} as a conservative choice. Instead, we consider the multi-wavelength fit in \citet{2026arXiv260111071T} as an independent analysis and explore how the use of a different data set affects the interpretation of DA 495’s physical properties (See Section \ref{sec:DA495Physics}). Lastly, consistent with this approach, we do not include flux points from the compact X-ray nebula, as we interpret it represents a physically distinct emission zone. 

To model the broadband emission, we use Naima, a Python package for analyzing non-thermal radiative emission from relativistic particles \citep{2015ICRC...34..922Z}. Since DA 495 is a PWN, which is powered by energetic electrons and positrons (hereafter electrons) \citep{2023PASA...40....7O}, we first test leptonic scenarios assuming the electron distributions described by PL and PLEC models with $E_0=1~\mathrm{TeV}$, and a broken power-law (BPL) model:
\begin{equation}
\frac{dN}{dE} = K 
\left\{
\begin{array}{ll}
\left( \frac{E}{E_b} \right)^{\alpha_1}, & E < E_b \\
\left( \frac{E}{E_b} \right)^{\alpha_2}, & E \geq E_b
\end{array}
\right..
\label{eq:Brokenpowerlaw}
\end{equation}
In the leptonic scenario, we consider synchrotron emission and inverse Compton scattering models, with target photon fields from the cosmic microwave background (CMB; $2.72~\mathrm{K},0.261~\mathrm{eV/cm^3}$), far-infrared (FIR; $30~\mathrm{K},0.5~\mathrm{eV/cm^3}$), and near-infrared (NIR; $3000~\mathrm{K},1~\mathrm{eV/cm^3}$) components, as implemented in Naima \citep{2010PhRvD..82d3002A, 2014ApJ...783..100K}. In the modeling, we fix the distance to the source to $1~\mathrm{kpc}$ \citep{2008ApJ...687..516K}, and fit the spectral parameters and the magnetic field strength.

\begin{figure*}[t]
\centering
\includegraphics[width=0.8\columnwidth]{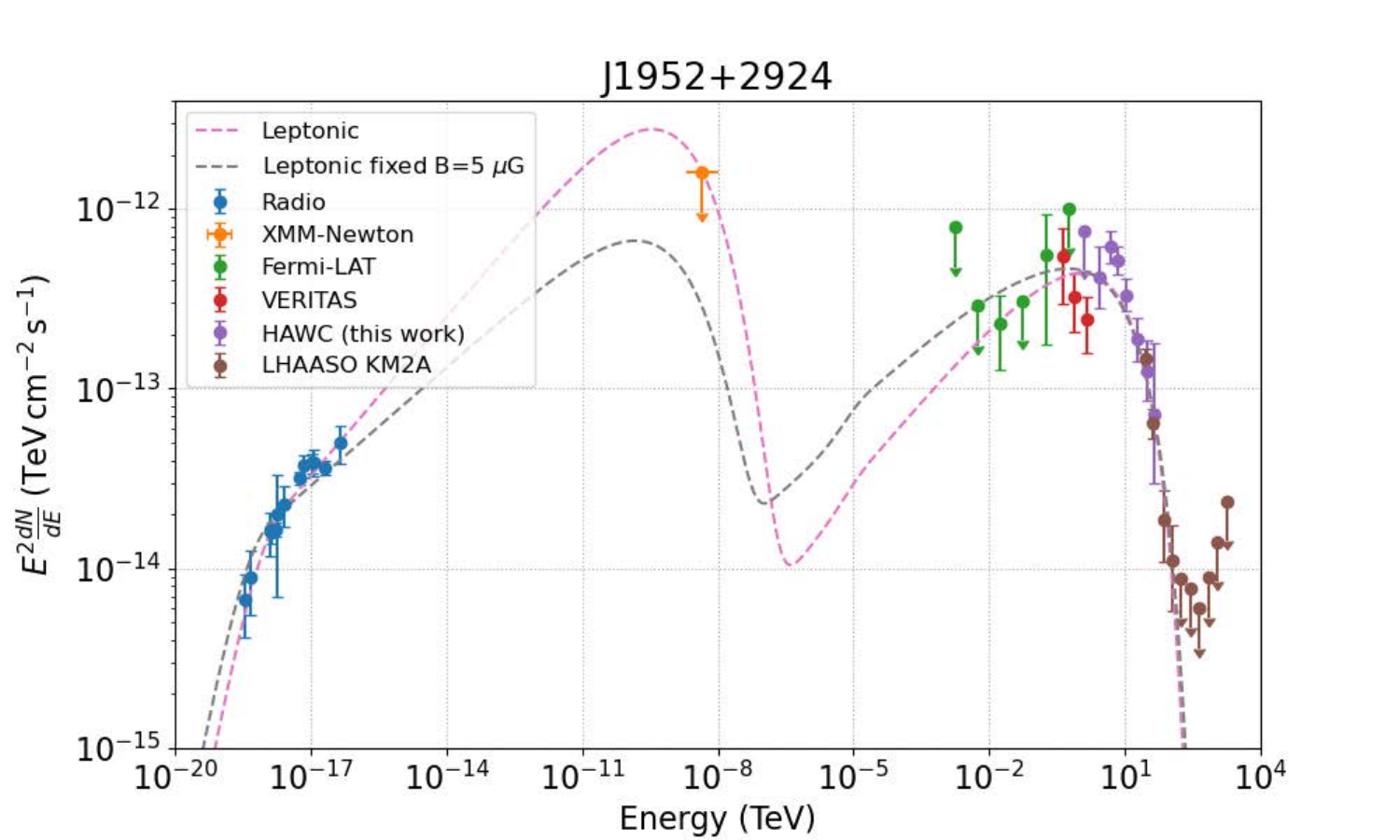}
\caption{SED modeling of J1952+2924 under a PLEC leptonic model. The pink dashed line represents a fit obtained by allowing all free parameters to vary, yielding a best-fit magnetic field of $B\sim10~\mu\mathrm{G}$. The gray dashed line shows an alternative fit in which the magnetic field is fixed at $5~\mu\mathrm{G}$. This constrained model is less consistent with the radio and GeV data, explaining why the unconstrained fit yields a higher magnetic field.}
\label{fig:J1952MWL}
\end{figure*}

Among the three distributions, the BPL model shows the minimum $\mathrm{AIC}$. However, it results in a posterior distribution of $\alpha_2$ ranging from $-8$ to $-4$ with a peak at $-5$, which we regard as unphysical. Such an extremely soft spectrum in the BPL model implies a sharp drop-off above the break energy, indicating a cutoff at higher energies. On the other hand, the PL model shows $\Delta \mathrm{AIC}>10$ compared to the BPL and PLEC models. In addition to leptonic models, we also perform modeling with hadronic components and find that they do not improve the fit. Therefore, the leptonic PLEC model is chosen as the best model, fitting $K,~\alpha,~E_c$ in Eq. (\ref{eq:power-law cutoff}) and the magnetic field $B$. Our fit yields a magnetic field of $B\sim10~\mu\mathrm{G}$, which is consistent with \citet{2019ApJ...878..126C} but higher than that from \citet{2026arXiv260111071T} ($\sim5~\mu\textrm{G}$) and \citet{2026ApJ...997..367T} ($\sim6~\mu\textrm{G}$). This discrepancy is mainly attributed to the use of different datasets. To account for the uncertainty in the magnetic field, we additionally fit the data by fixing the magnetic field to be $5~\mu\textrm{G}$ and parameter shifts in the other three parameters are considered systematic uncertainties, confirming that they are stable under different magnetic fields. Figure \ref{fig:J1952MWL} presents the multi-wavelength fit of DA 495, and Table \ref{tab:J1952MWL} shows the fit results.



\begin{deluxetable*}{cccccc}
\tablewidth{0pt}
\tablecaption{Leptonic PLEC modeling results for J1952}
\label{tab:J1952MWL}
\tablehead{
\colhead{Source} & \colhead{$\log{K}$} & \colhead{$\alpha$} & \colhead{$\log{E_{cut}}$} & \colhead{$\log{B}$} & 
}
\startdata
J1952 &
$44.83\,^{+0.04}_{-0.04}\,^{+0.15}~(44.85)$ & $2.42\,^{+0.03}_{-0.05}\,^{+0.15}~(2.41)$ & $1.70\,^{+0.05}_{-0.05}\,^{+0.08}~(1.69)$ & $0.99\,^{+0.10}_{-0.04}~(1.00)$ \\
\enddata
\tablecomments{The reported values correspond to the posterior median with the maximum-likelihood estimates given in parentheses. The units of $K, E_{cut}$, and $B$ are $\text{TeV}^{-1}$, $\text{TeV}$, and $\mu\text{G}$, respectively. The first uncertainty is statistical, and the second is a one-sided systematic from the fit with fixed $B=5~\mu\textrm{G}$. Since the systematic uncertainties are derived by fixing $B=5~\mu\textrm{G}$, no systematic uncertainty is reported for $B$ itself.}
\end{deluxetable*}

\begin{deluxetable*}{ccccc}
\tablewidth{0pt}
\tablecaption{Leptonic modeling results for J1954 and J1958}
\label{tab:J1954J1958}
\tablehead{
\colhead{Source} & \colhead{$\log{K}$} & \colhead{$\alpha$} & \colhead{$\log{E_{cut}}$} &
}
\startdata
J1954 &
$44.60\,^{+1.18}_{-1.98}~(45.10)$ & $1.71\,^{+1.01}_{-1.74}~(2.13)$ & $1.54\,^{+0.57}_{-0.37}~(1.70)$ \\
J1958 &
$45.13\,^{+0.93}_{-2.31}~(44.76)$ & $2.27\,^{+0.63}_{-1.64}~(1.97)$ & $2.08\,^{+0.99}_{-0.52}~(1.90)$ \\
\enddata
\tablecomments{The reported values correspond to the posterior median with asymmetric upper and lower bounds. Values in parentheses indicate the maximum-likelihood estimates. The units of $K$, and $E_{cut}$ are $\text{TeV}^{-1}$ and $\text{TeV}$. The uncertainties are statistical.}
\end{deluxetable*}

\begin{figure}[t]
\centering
\includegraphics[width=0.7\columnwidth]{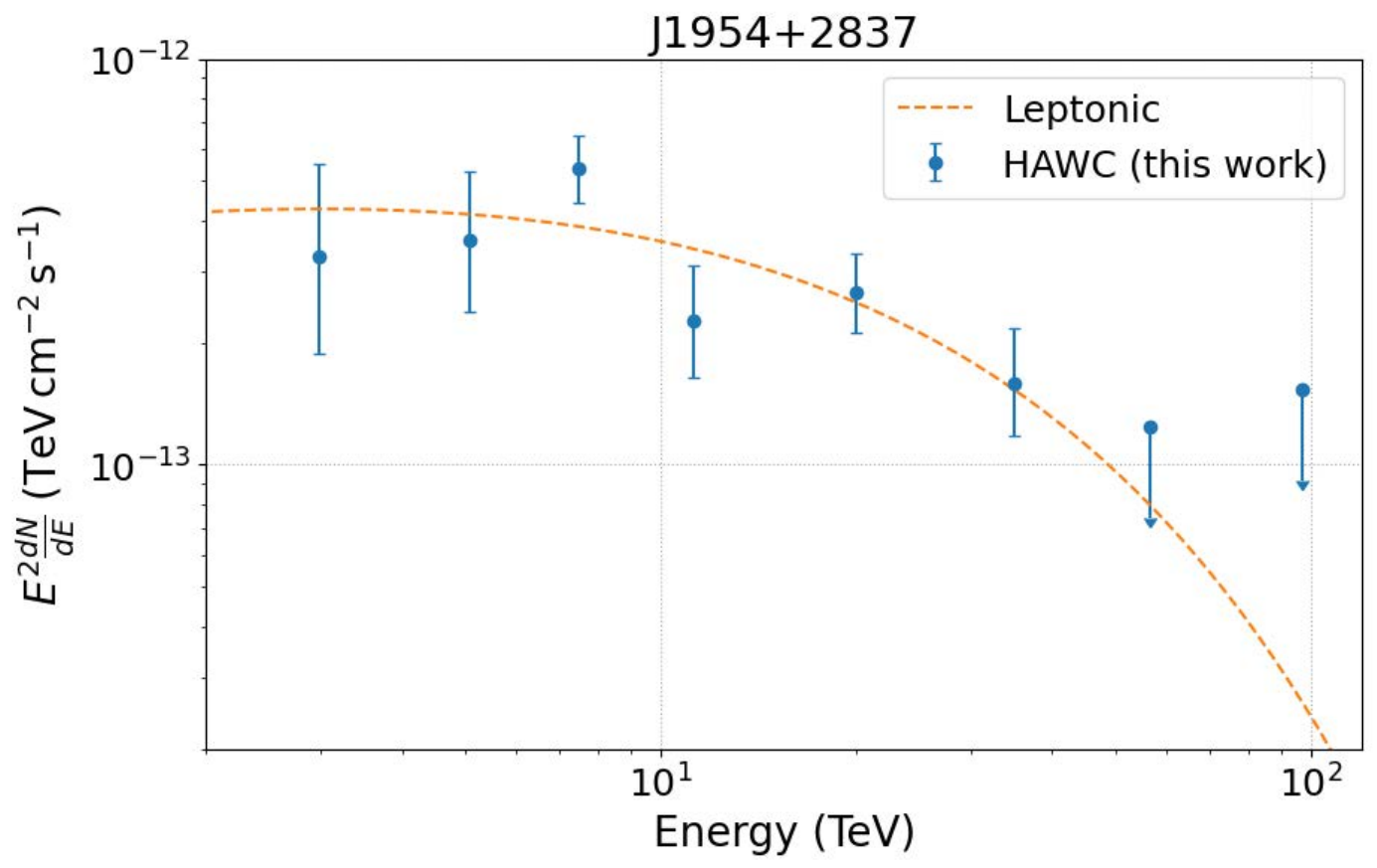}
\caption{SED modeling of J1954+2837 under a PLEC leptonic model.}
\label{fig:J1954Model}
\end{figure}

\begin{figure}[t]
\centering
\includegraphics[width=0.7\columnwidth]{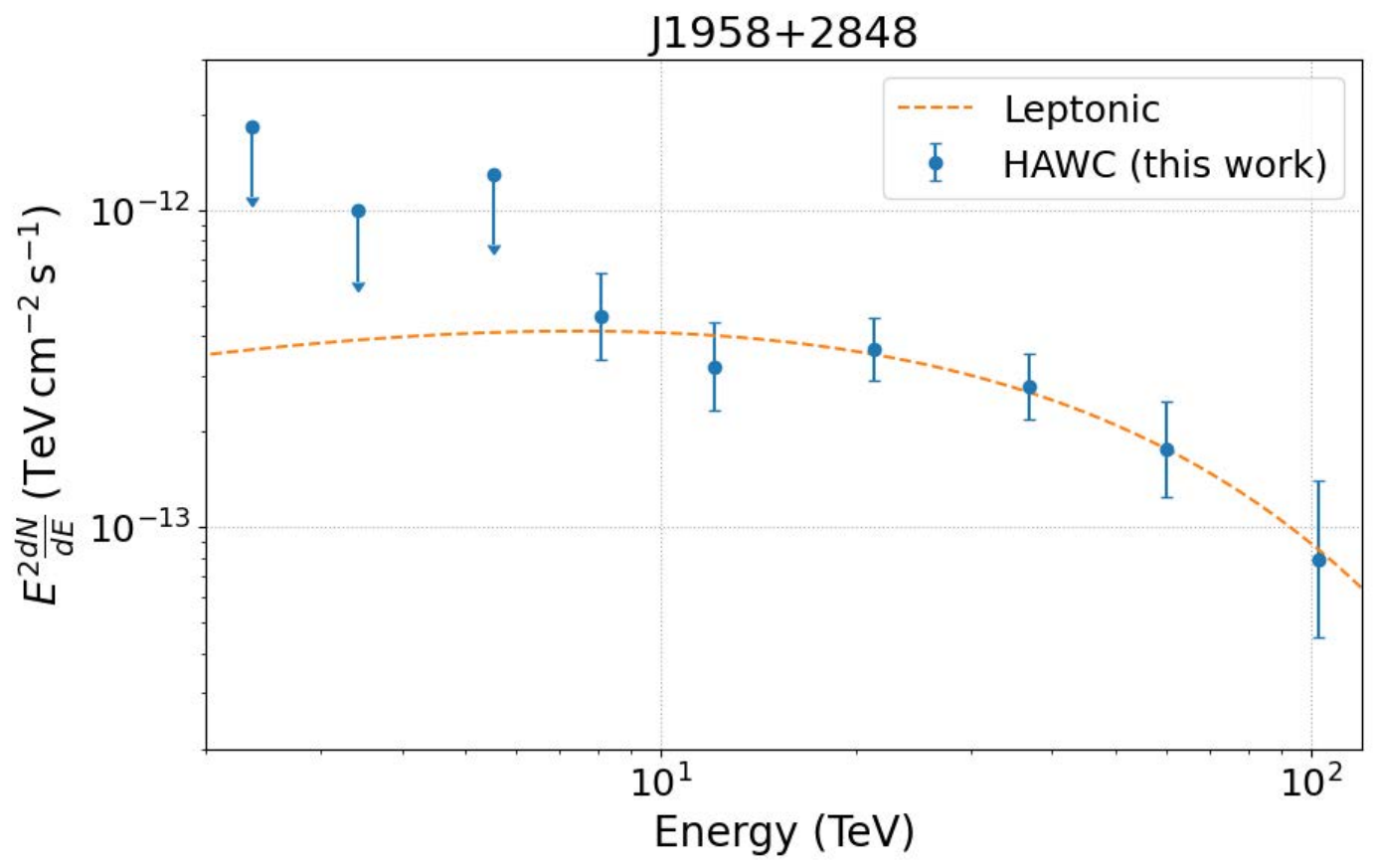}
\caption{SED modeling of J1958+2848 under a PLEC leptonic model.}
\label{fig:J1958Model}
\end{figure}

\subsection{HAWC J1954+2837 and HAWC J1958+2848}
\label{sec:SourceModeling2}
HAWC J1954+2837 and HAWC J1958+2848 remain unidentified, with no detections at other wavelengths. However, \citet{2022ApJ...930..164X} and \citet{2023ApJ...957...79Z} suggested that they are TeV halos associated with nearby middle-aged pulsars PSR J1954+2836 \citep{2010ApJ...725..571S} and PSR J1958+2846 \citep{2009Sci...325..840A}, respectively. To assess this hypothesis, we model them using the three leptonic models, PL, PLEC, and BPL, with Naima as described in Section \ref{sec:DA495Modeling}, since TeV halos are characterized by diffuse electrons \citep{PhysRevD.100.043016}, and adopt distances of $1.96~\textrm{kpc}$ for J1954 and $1.95~\textrm{kpc}$ for J1958 from version 2.8.0 of the Australia Telescope National Facility (ATNF) pulsar catalog \citep{2005AJ....129.1993M}. In the leptonic scenario, we consider only inverse Compton scattering with CMB, FIR, and NIR photon fields, as this process dominates in the VHE range. For both sources, we find that the BPL model yields $\Delta\mathrm{AIC}>2$ compared to the PL model, which shows the lowest $\mathrm{AIC}$. However, despite this statistical preference, the PL model requires an unrealistic electron efficiency. On the other hand, even though the PLEC model yields a slightly higher $\mathrm{AIC}$ than the PL model ($\Delta\mathrm{AIC}<2$), it provides a more physically plausible scenario (see Section \ref{sec:TwoTeVHalos}). Therefore, we select the PLEC model to account for the VHE gamma-ray emission from these sources. Figures \ref{fig:J1954Model} and \ref{fig:J1958Model} show modeling results for J1954 and J1958, summarized in Table \ref{tab:J1954J1958}.

\section{Discussion}
\label{sec:Discussion}

\subsection{Potential Fourth Source?}
\label{subsec:FourthSource}
Although we mainly focus on the properties of the three sources, J1952, J1954, and J1958, our analysis suggests the presence of a potential fourth source in the DA 495 region, characterized by a large spatial extension. When $k_{GDE}$ is allowed to vary in the morphological analysis, adding a fourth extended source yields a TS value close to our detection threshold. Moreover, four extended sources are found when $k_{GDE}$ is fixed. Table~\ref{tab:FourthSource} compares the properties of the potential fourth source under different GDE modeling assumptions. The uncertainties are large, and there are significant discrepancies in the flux, indicating that its characteristics depend strongly on the GDE model. Therefore, we consider it premature to claim it as a distinct source, though potential counterparts include SNRs G066.0-0.0 and G067.6+0.9 \citep{2013MNRAS.431..279S}, and a possible TeV halo associated with PSR J1958+3033 \citep{2018ApJ...869..181P}. Future observations, improved GDE models, and more precise reconstruction tools will enable a detailed investigation of this source candidate.

\begin{deluxetable*}{cccccc}
\tablewidth{0pt}
\tablecaption{Comparison of the potential fourth source for free and fixed GDE normalization fits}
\label{tab:FourthSource}
\tablehead{
\colhead{R.A. (deg)} & \colhead{Decl. (deg)} & \colhead{Extension (deg)} & \colhead{$K~(\mathrm{TeV}^{-1}\mathrm{cm}^{-2}\mathrm{s}^{-1})$} & \colhead{$\alpha$} & \colhead{$k_{GDE}$}
}
\startdata
$299.36\pm0.23$ & $30.36\pm0.29$ & $0.64\pm0.22$ & $7.98^{+5.79}_{-3.25}\times10^{-14}$ & $-2.44\pm0.10$ & $3.3\pm0.7$\\
$299.21\pm0.19$ & $29.80\pm0.47$ & $1.15\pm0.29$ & $26.89^{+8.02}_{-6.45}\times10^{-14}$ & $-2.51\pm0.06$ & $1~\text{(fixed)}$ \\
\enddata
\tablecomments{Only statistical uncertainties are reported. Spectral parameters are obtained using the PL model with fixed $E_0=7~\textrm{TeV}$.}
\end{deluxetable*}


\begin{figure}[t]
\centering
\includegraphics[width=0.6\columnwidth]{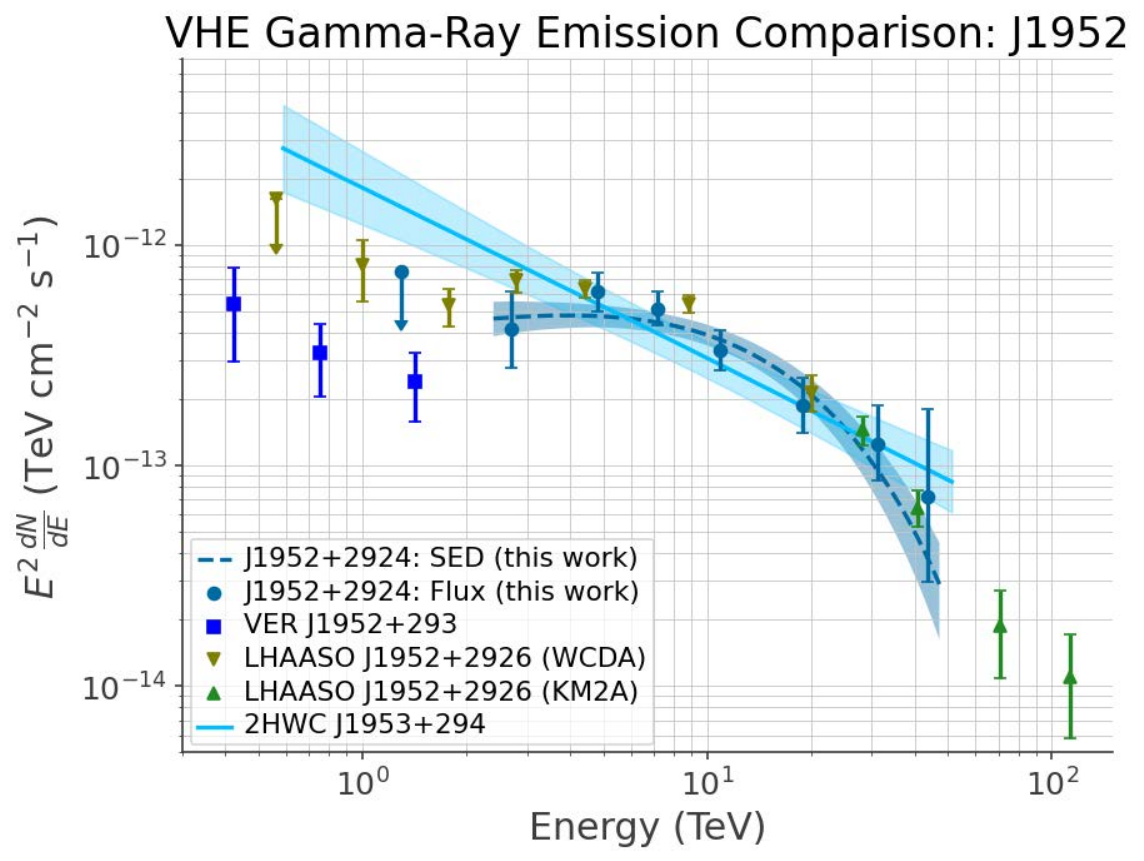}
\caption{Spectral comparison of J1952 with VER J1952+293 and LHAASO J1952+2926. The VERITAS flux points are taken from \citet{2018ApJ...866...24A}, the LHAASO flux points are taken from \citet{2026arXiv260111071T}, and the 2HWC spectrum is reproduced following \citet{2017ApJ...843...40A}.}
\label{fig:J1952Comparison}
\end{figure}
\begin{figure}[t]
\centering
\includegraphics[width=0.6\columnwidth]{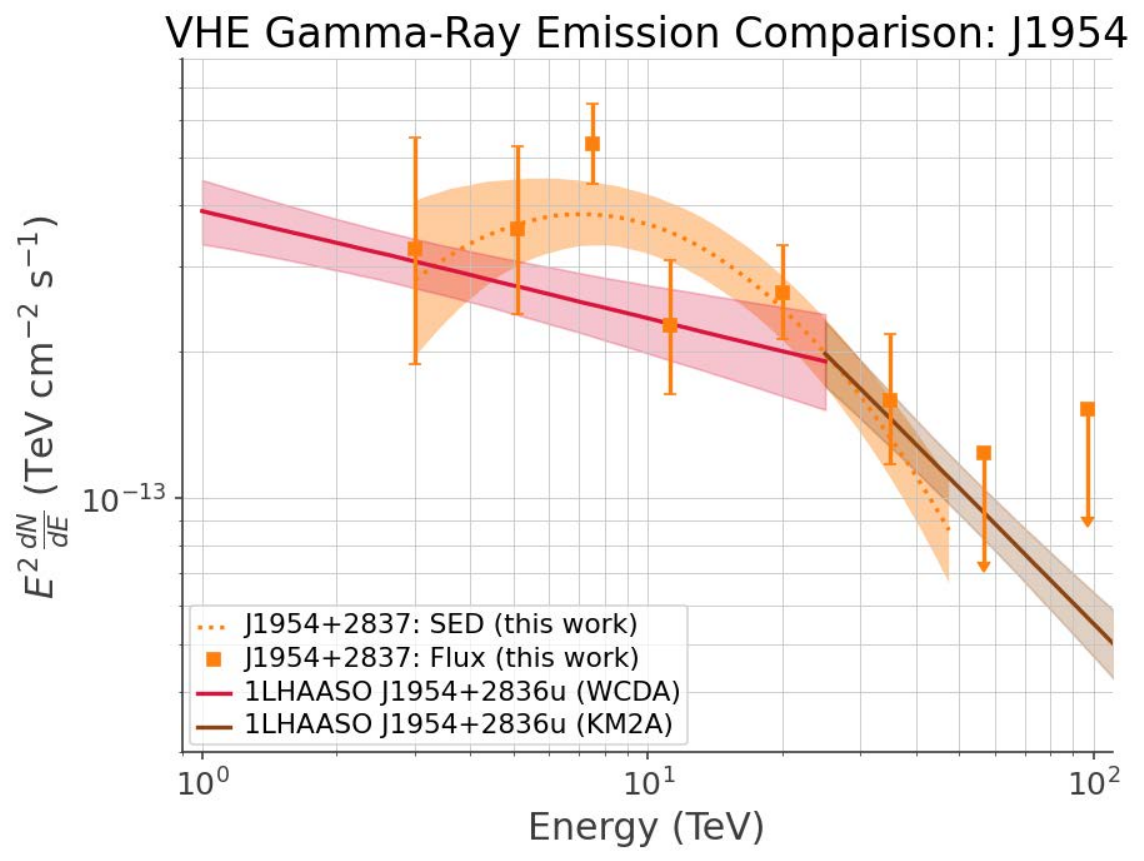}
\caption{Spectral comparison of J1954 with 1LHAASO J1954+2836u. The LHAASO spectra are reproduced following \citet{2024ApJS..271...25C}.}
\label{fig:J1954Comparison}
\end{figure}
\begin{figure}[t]
\centering
\includegraphics[width=0.6\columnwidth]{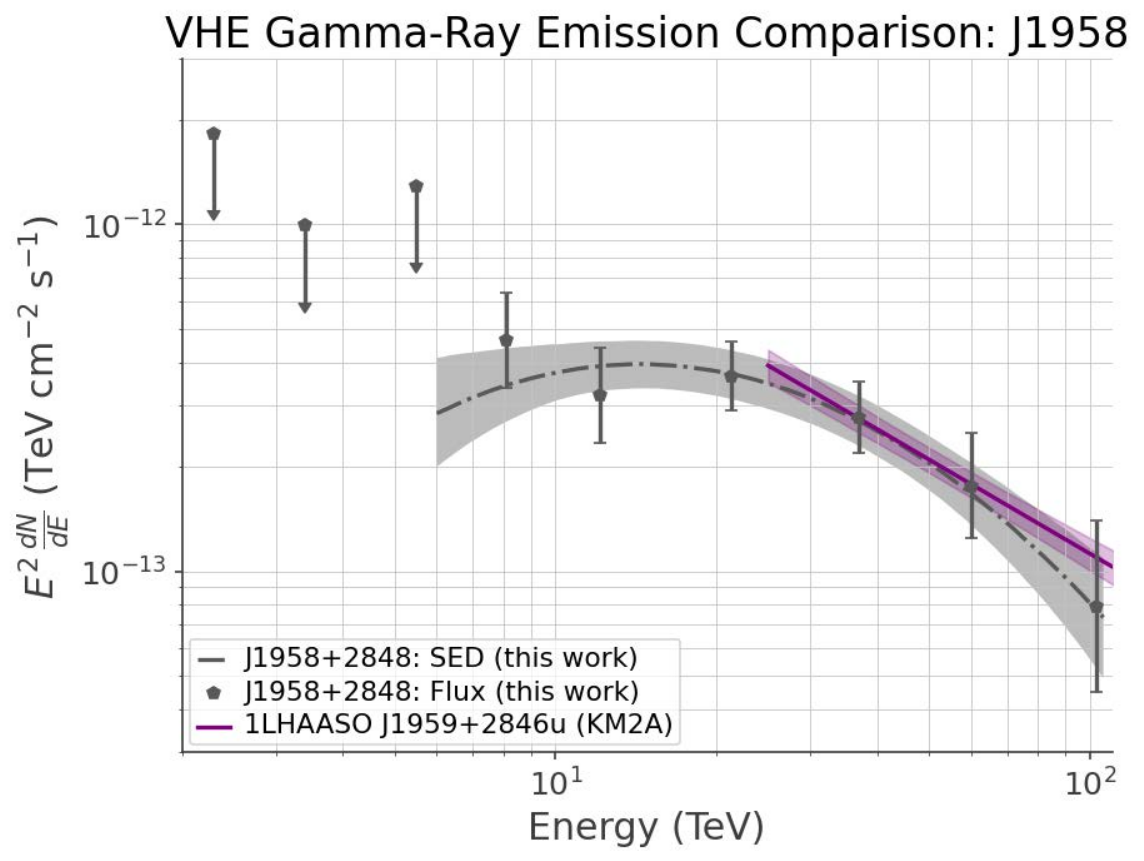}
\caption{Spectral comparison of J1958 with 1LHAASO J1959+2846u. The LHAASO spectrum is reproduced following \citet{2024ApJS..271...25C}.}
\label{fig:J1958Comparison}
\end{figure}

\subsection{Spectral Comparison with Other Experiments}
The flux discrepancy for DA 495 between HAWC and VERITAS has persisted since the analysis presented by \citet{2018ApJ...866...24A}. The discrepancy at 1 TeV was about a factor of 7 between 2HWC and VERITAS, and it decreased to a factor of 3 to 4 in 3HWC and 4HWC. Our new spectral results significantly reduce this tension. The differential flux at 1 TeV measured by VERITAS was $(2.84\pm0.54_{\mathrm{stat}})\times10^{-13}~ \mathrm{TeV}^{-1}\mathrm{cm}^{-2}\mathrm{s}^{-1}$, while that from our best-fit spectrum extrapolated to 1 TeV is $(3.89\pm1.27_{\mathrm{stat}})\times10^{-13}~ \mathrm{TeV}^{-1}\mathrm{cm}^{-2}\mathrm{s}^{-1}$, which are consistent within a $1\sigma$ level. This improvement mainly results from the detailed spectral analysis that allows us to probe spectral features in the few-TeV energy range, with higher statistics and improved reconstruction methods playing a secondary role. Additionally, we compared our spectral results with those from LHAASO, using the joint WCDA and KM2A analysis results for DA 495 and the 1LHAASO catalog for the other two sources, J1954 and J1958. The agreement shown in Figures \ref{fig:J1952Comparison}, \ref{fig:J1954Comparison}, and \ref{fig:J1958Comparison} demonstrates consistency across independent experiments. 

\begin{deluxetable*}{lccccccccc}
\tablewidth{0pt}
\tablecaption{Properties of the TeV halo candidates J1954 and J1958}
\label{tab:TeVHalos}
\tablehead{
\colhead{Source} & 
\colhead{$\tau_c~\textrm{(kyr)}$} &
\colhead{$\Delta\theta~\textrm{(deg)}$} & 
\colhead{$L_\gamma~\textrm{(erg/s)}$} & 
\colhead{$\dot{E}~\textrm{(erg/s)}$} & 
\colhead{$\epsilon_\gamma~(\%)$} &
\colhead{$W_e~\textrm{(erg)}$}& 
\colhead{$\eta_e~(\%)$} & 
\colhead{$D~(\mathrm{cm}^{2}~\mathrm{s}^{-1})$} &
\colhead{$D_{ISM}~(\mathrm{cm}^{2}~\mathrm{s}^{-1})$}
}
\startdata
J1954 & $69.4~\mathrm{kyr}$ & $0.07^\circ$ & $5.8\times10^{32}$ & 
$1.0\times10^{36}$ & $0.06$ & 
$2.6\times10^{46}$ & $1$ & 
{$3.2\times10^{26}$} &
{$1.0\times10^{30}$} \\
J1958 & $21.7~\mathrm{kyr}$ & $0.09^\circ$ & $5.9\times10^{32}$ & 
$3.4\times10^{35}$ & $0.17$ & 
$9.5\times10^{45}$ & $4$ & 
$1.4\times10^{27}$ &
$1.2\times10^{30}$\\
\enddata
\tablecomments{Pulsar properties are adopted from version 2.8.0 of the ATNF pulsar catalog. The diffusion coefficients are evaluated at the best-fit cutoff energies ($E_c=50~\textrm{and}~79~\mathrm{TeV}$). For comparison, we include the diffusion coefficient in the general ISM, $D_{ISM}$, estimated by $D_{ISM}=D_0(E_c/3~\mathrm{GeV})^{1/3}$ with $D_{0}=4\times10^{28}~\mathrm{cm}^{2}~\mathrm{s}^{-1}$}, following \cite{2016ApJ...824...16J}.
\end{deluxetable*}

\subsection{PWN DA 495}
\label{sec:DA495Physics}
Although DA 495 was first detected decades ago, many of its properties remain unclear, mainly due to the absence of a detected central pulsar powering the nebula. For instance, \citet{2008ApJ...687..516K} estimated the distance to DA 495 to be $1.0\pm0.4~\mathrm{kpc}$ based on HI absorption measurements, whereas \citet{2015MNRAS.453.2241K} suggested distances of $2.4^{+1.3}_{-1.1}, 3.3^{+1.7}_{-1.3},~\mathrm{or}~3.2^{+1.7}_{-1.3}~\mathrm{kpc}$ from the interstellar absorption-distance relation, depending on different models. In addition, the age of DA 495 was estimated to be $20~\mathrm{kyr}$ in \citet{2008ApJ...687..516K} by comparing its physical properties with those of other PWNe. However, \citet{2015MNRAS.453.2241K} reported a wide range of $7$ to $155~\mathrm{kyr}$ and different estimates from \citet{2019ApJ...878..126C} yielded $3.5\pm0.5~\mathrm{kyr}$ and $>1~\mathrm{Myr}$, suggesting DA 495 may not follow a typical PWN evolutionary path. Among the several estimates, we use a distance of $1~\textrm{kpc}$ and an age of $20~\textrm{kyr}$ as reference values in the following discussion. Since DA 495 is considered a middle-aged PWN under this age assumption, we do not distinguish between its age and the characteristic age of the putative pulsar.

Our new spectral results and modeling of DA 495 enable a detailed investigation of the physical properties of this puzzling source. In the following calculations, we assume a cutoff energy of $50~\mathrm{TeV}$ for the electron population and a magnetic field strength of $10~\mu\mathrm{G}$, as obtained from the modeling described in Section \ref{sec:DA495Modeling}. We also consider an alternative scenario with $B=5~\mu\mathrm{G}$ for a more thorough investigation. Using these assumptions, we first estimate the cooling time $t_{cool}$ to explore the origin of the spectral cutoff in the electron distribution, which is calculated as
\begin{equation}
t_{cool}=3.1\times10^5\left(\frac{1}{E_{e}~(\mathrm{TeV})~U_{tot}~(\mathrm{eV/cm^3})}\right)~\mathrm{yr},
\label{eq:coolingtime}
\end{equation}
where $E_e$ represents the energy of electrons and $U_{tot}$ is the sum of the magnetic field energy density and the energy density of CMB photons. The contributions from FIR and NIR photon fields are neglected, as the interactions between electrons and these photons are highly suppressed, barely affecting the result of the calculation of $t_{cool}$. For $50~\mathrm{TeV}$ electrons under a magnetic field of $10~\mu\mathrm{G}$, we find the cooling time $t_{cool}\sim2.3~\mathrm{kyr}$, which is shorter than the suggested age candidates of DA 495. Additionally, when $B=5~\mu\textrm{G}$, we obtain $t_{cool}\sim7.0~\mathrm{kyr}$, which still remains shorter than most of its age estimates. These results suggest that the spectral cutoff in the electron distribution originates from 
radiative cooling, dominated by synchrotron losses, rather than being intrinsic to the acceleration mechanism.

Based on the calculations of $t_{cool}$, we estimate the diffusion coefficient $D$ as
\begin{equation}
D=\frac{d_{dif}^2}{4t_{cool}},
\label{eq:DiffusionCoefficient}
\end{equation}
where $d_{dif}$ is a diffusion radius. Assuming a distance to the source of $1~\mathrm{kpc}$ and a spatial extension of $0.12^\circ$, we obtain $d_{dif}\sim2~\mathrm{pc}$, yielding $D\sim1.5\times10^{26}~\mathrm{cm}^2\mathrm{s}^{-1}$. Comparing this with the Bohm diffusion limit $D_{Bohm}\sim1.7\times10^{26}~\mathrm{cm}^2\mathrm{s}^{-1}$, given by
\begin{equation}
D_{Bohm}=3.34\times10^{19}\frac{E_e~(\mathrm{TeV})}{B~(\mathrm{G})}~\mathrm{cm}^2\mathrm{s}^{-1},
\label{eq:BohmDiffusion}
\end{equation}
assuming $B=10~\mu\mathrm{G}$, we find $D$ smaller than $D_{Bohm}$ (for the derivation of Eq. (\ref{eq:BohmDiffusion}), see \citet{1983RPPh...46..973D}). Since $D_{Bohm}$ represents the minimum possible diffusion coefficient, $D<D_{Bohm}$ is non-physical. It can be resolved by assuming a larger distance of $1.4~\mathrm{kpc}$, considering the uncertainty in the measurement $(0.4~\mathrm{kpc})$. However, the corresponding $D$ value is $2.9\times10^{26}~\mathrm{cm}^2\mathrm{s}^{-1}$, which remains close to $D_{Bohm}$. For $B=5~\mu\mathrm{G}$, the inconsistency becomes more severe, as a lower magnetic field decreases $D$ (through a longer $t_{cool}$) and increases $D_{Bohm}$, widening the gap between the two parameters. These results may indicate that DA 495 is located at a larger distance, as proposed by \citet{2015MNRAS.453.2241K}. When $B=10~\mu\mathrm{G}$, their distance estimates of $2.4-3.3~\textrm{kpc}$ yield $D\sim10^{27}~\mathrm{cm}^2\mathrm{s}^{-1}$, which is safely above the Bohm diffusion limit. On the other hand, when $B=5~\mu\mathrm{G}$, a distance of $\sim3~\mathrm{kpc}$ or beyond is required to satisfy $D>D_{Bohm}$, consistently supporting a larger distance to the source.

We further investigate the distance and age of DA 495 using the PWN population study conducted by the High Energy Stereoscopic System (H.E.S.S.) collaboration~\citep{2018A&A...612A...2H}, which provides the relation between the luminosity of a PWN from 1 to 10 TeV, $L_{1-10~\mathrm{TeV}}$, and the characteristic age of its associated pulsar, $\tau_{c}$. \citet{2019ApJ...878..126C} previously reported $L_{1-10~\mathrm{TeV}}$ of $(1.3\pm0.2)\times10^{31}~\mathrm{erg}~\mathrm{s}^{-1}$, estimating an age of $>$1 Myr assuming $d=1~\mathrm{kpc}$. We compute $L_{1-10~\mathrm{TeV}}$ using our spectral results through the gamma-ray luminosity $L_{\gamma}$, calculated as
\begin{equation}
L_\gamma=4\pi d^2 \int_{E_{min}}^{E_{max}}E\frac{dN}{dE}dE,
\label{eq:Luminosity}
\end{equation}
where $d$ is the distance to the source. By setting $E_{min}=1~\mathrm{TeV}$, $E_{max}=10~\mathrm{TeV}$, and assuming  $d=1~\mathrm{kpc}$, we obtain $L_{1-10~\mathrm{TeV}}=(2.0\pm1.6)\times10^{32}~\mathrm{erg}~\mathrm{s}^{-1}$, which now places DA 495 within the PWN population band for an age of $20~\mathrm{kyr}$, making the combination of its age and distance consistent with the population study. Notably, for a fixed age of $20~\mathrm{kyr}$, increasing the distance shifts the source closer to the main population fit, as does increasing the age for a fixed distance of $1~\mathrm{kpc}$. This may indicate that DA 495 is located beyond $1~\mathrm{kpc}$ and/or older than $20~\mathrm{kyr}$.

\subsection{Two TeV Halo Candidates}
\label{sec:TwoTeVHalos}
To assess the TeV halo candidacy of J1954 and J1958, we estimate the angular separations from their nearby pulsars, the gamma-ray and electron efficiencies, and the diffusion coefficient. The angular separations $\Delta\theta$ from PSR J1954+2836 and PSR J1958+2846 are $0.07^\circ$ and $0.09^\circ$ for J1954 and J1958, respectively. The gamma-ray efficiency,
\begin{equation}
\epsilon_\gamma=\frac{L_\gamma}{\dot{E}},
\label{eq:Gamma-RayEfficiency}
\end{equation}
is calculated using $L_{\gamma}$ and the spin-down power of a pulsar $\dot{E}$, yielding 0.06$\%$ and 0.17$\%$ for J1954 and J1958, consistent with the results from TeV halo stacking analysis \citep{2025PhRvL.134q1005A}. Additionally, the electron efficiency is defined as
\begin{equation}
\eta_e=\frac{W_e}{\tau_{c}\dot{E}},
\label{eq:ElectronEfficiency}
\end{equation}
where $W_e$ is the total energy in electrons derived from the results in Section \ref{sec:SourceModeling2}. Assuming the leptonic PLEC distribution, we obtain 1$\%$ and $4\%$ for J1954 and J1958, which are consistent with the few-percent efficiencies derived from other known TeV halos \citep{2019PhRvD.100l3015D, 2022PhRvD.106l3017F, 2024ApJ...974..246A}. Lastly, the diffusion coefficient is estimated using Eq. (\ref{eq:coolingtime}) and (\ref{eq:DiffusionCoefficient}) for electrons at $E_{c}$, assuming a typical magnetic field of $5~\mu\mathrm{G}$ in the ISM \citep{2015ASSL..407..483H}, which yields values that are $\gtrsim1000$ times smaller than those in the ISM. These results, summarized in Table \ref{tab:TeVHalos}, support the interpretation that J1954 and J1958 are TeV halos associated with nearby pulsars.

\section{Conclusion}
\label{sec:Conclusion}
We perform a dedicated analysis of the DA 495 region using 3071 days of HAWC data. In the morphological analysis, we identify three extended sources, HAWC J1952+2924 (DA 495), HAWC J1954+2837, and HAWC J1958+2848, whose fit parameters are robust 
under variations in the GDE normalization assumption. Additionally, we find a hint of a fourth extended source, but its properties are highly sensitive to this assumption. Future observations and refined GDE models will be necessary to explore the potential fourth source. Our spectral analysis reveals curved VHE gamma-ray spectra for all three sources, with the PLEC model for HAWC J1952+2924 and the LP model for HAWC J1954+2837 and HAWC J1958+2848. These measurements resolve the flux discrepancy of DA 495 between HAWC and VERITAS and show consistency between HAWC and LHAASO for the three sources. Using these morphological and spectral results, we investigate DA 495 and the other two sources. Our best-fit leptonic PLEC model for DA 495 yields a cooling time shorter than its age, suggesting that the spectral cutoff in the electron distribution originates from the cooling processes of the electron population. Moreover, the estimate of the diffusion coefficient and the PWN population study support the interpretation that DA 495 may be located beyond $1~\mathrm{kpc}$ and older than $20~\mathrm{kyr}$. Additionally, the origin of the TeV gamma-ray emission from the other two sources is likely to be TeV halos associated with nearby pulsars PSR J1954+2836 and PSR J1958+2846, supported by their angular separations, as well as the estimates of the gamma-ray and electron efficiencies, and the diffusion coefficient. 


\begin{acknowledgments}
We acknowledge the support from: the US National Science Foundation (NSF); the US Department of Energy Office of High-Energy Physics; the Laboratory Directed Research and Development (LDRD) program of Los Alamos National Laboratory; Consejo Nacional de Ciencia y Tecnolog\'{i}a (CONACyT), M\'{e}xico, grants LNC-2023-117, 271051, 232656, 260378, 179588, 254964, 258865, 243290, 132197, A1-S-46288, A1-S-22784, CF-2023-I-645, CBF2023-2024-1630, c\'{a}tedras 873, 1563, 341, 323, Red HAWC, M\'{e}xico; DGAPA-UNAM grants IG101323, IG100726, IN111716-3, IN111419, IA102019, IN106521, IN114924, IN110521, IN102223; VIEP-BUAP; PIFI 2012, 2013, PROFOCIE 2014, 2015; the University of Wisconsin Alumni Research Foundation; the Institute of Geophysics, Planetary Physics, and Signatures at Los Alamos National Laboratory; Polish Science Centre grant, 2024/53/B/ST9/02671; Coordinaci\'{o}n de la Investigaci\'{o}n Cient\'{i}fica de la Universidad Michoacana; Royal Society - Newton Advanced Fellowship 180385; Gobierno de Espa\~{n}a and European Union-NextGenerationEU, grant CNS2023-144099; The Program Management Unit for Human Resources \& Institutional Development, Research and Innovation, NXPO (grant number B16F630069); Coordinaci\'{o}n General Acad\'{e}mica e Innovaci\'{o}n (CGAI-UdeG), PRODEP-SEP UDG-CA-499; Institute of Cosmic Ray Research (ICRR), University of Tokyo. H.M. acknowledges support under grant number CBF2023-2024-1630. H.F. acknowledges support by NASA under award number 80GSFC21M0002. C.R. acknowledges support from National Research Foundation of Korea (RS-2023-00280210). We also acknowledge the significant contributions over many years of Stefan Westerhoff, Gaurang Yodh and Arnulfo Zepeda Dom\'{i}nguez, all deceased members of the HAWC collaboration. Thanks to Scott Delay, Luciano D\'{i}az and Eduardo Murrieta for technical support.
\end{acknowledgments}

\appendix

\section{SED Flux Points}
\label{app:fluxpoints}

In Tables \ref{tab:J1952_SED}, \ref{tab:J1954_SED}, and \ref{tab:J1958_SED}, we provide the flux points used in Figure \ref{fig:Spectra}. Upper limits (ULs) are reported at the $90~\%$ confidence level.

\begin{table}[htb!]
\centering
\caption{SED flux points for J1952}
\label{tab:J1952_SED}
\begin{tabular}{rc}
\hline \hline
$E$ (TeV) & $E^2~dN/dE$ ($10^{-13}~\mathrm{TeV}~\mathrm{cm}^{-2}~\mathrm{s}^{-1}$) \\
\hline
1.3  & $7.59~\mathrm{(UL)}$ \\
2.7  & $4.17^{+2.00}_{-1.38}$ \\
4.8  & $6.16^{+1.37}_{-1.15}$ \\
7.2  & $5.17^{+0.96}_{-0.82}$ \\
10.9 & $3.31^{+0.79}_{-0.62}$ \\
18.9 & $1.88^{+0.61}_{-0.47}$ \\
31.2 & $1.26^{+0.61}_{-0.40}$ \\
43.7 & $0.72^{+1.09}_{-0.42}$ \\
\hline
\end{tabular}
\tablecomments{Only statistical uncertainties are reported.}
\end{table}

\begin{table}[htb!]
\centering
\caption{SED flux points for J1954}
\label{tab:J1954_SED}
\begin{tabular}{rc}
\hline \hline
$E$ (TeV) & $E^2dN/dE$ ($10^{-13}~\mathrm{TeV}~\mathrm{cm}^{-2}~\mathrm{s}^{-1}$) \\
\hline
3.0  & $3.26^{+2.25}_{-1.38}$ \\
5.1  & $3.57^{+1.72}_{-1.19}$ \\
7.5  & $5.36^{+1.12}_{-0.93}$ \\
11.2 & $2.27^{+0.82}_{-0.63}$ \\
20.0 & $2.65^{+0.67}_{-0.52}$ \\
34.9 & $1.59^{+0.58}_{-0.41}$ \\
56.7 & $1.24~\mathrm{(UL)}$ \\
96.8 & $1.53~\mathrm{(UL)}$ \\
\hline
\end{tabular}
\tablecomments{Only statistical uncertainties are reported.}
\end{table}

\begin{table}[htb!]
\centering
\caption{SED flux points for J1958}
\label{tab:J1958_SED}
\begin{tabular}{rc}
\hline \hline
$E$ (TeV) & $E^2dN/dE$ ($10^{-13}~\mathrm{TeV}~\mathrm{cm}^{-2}~\mathrm{s}^{-1}$) \\
\hline
2.3   & $18.27~\mathrm{(UL)}$ \\
3.4   & $9.94~\mathrm{(UL)}$ \\
5.5   & $12.90~\mathrm{(UL)}$ \\
8.1   & $4.64^{+1.69}_{-1.26}$ \\
12.1  & $3.21^{+1.21}_{-0.87}$ \\
21.3  & $3.65^{+0.93}_{-0.74}$ \\
36.9  & $2.77^{+0.76}_{-0.58}$ \\
60.0  & $1.76^{+0.74}_{-0.51}$ \\
102.9 & $0.79^{+0.61}_{-0.34}$ \\
\hline
\end{tabular}
\tablecomments{Only statistical uncertainties are reported.}
\end{table}


\clearpage

\bibliography{Main}{}
\bibliographystyle{aasjournalv7}



\end{document}